\documentclass[letter,11pt]{article}
\pdfoutput=1 % if your are submitting a pdflatex (i.e. if you have
\usepackage{jcappub} % for details on the use of the package, please
\usepackage[T1]{fontenc} % if needed

\usepackage{amssymb}
\usepackage{graphicx, epsfig, bm, amsmath}
\usepackage{hyperref}

\definecolor{darkgreen}{cmyk}{0.85,0.2,1.00,0.2}
\usepackage{color}
\usepackage{ifthen}
\def\nn{\nonumber}
\def\({\left(}
\def\){\right)}
\def\[{\left[}
\def\]{\right]}
\def\tr{{\rm Tr}}
\def\gfc{\xi}
\newcommand{\beq}{\begin{equation}}
\newcommand{\beqn}{\begin{eqnarray}}
\newcommand{\eeq}{\end{equation}}
\newcommand{\eeqn}{\end{eqnarray}}

\bibstyle{JHEP}

\title{Gauge-preheating and the end of axion inflation}

\author[a]{Peter Adshead}
\author[b,c]{John T. Giblin, Jr.}
\author[b]{Timothy R. Scully}
\author[a]{Evangelos I. Sfakianakis}

\affiliation[a]{Department of Physics, University of Illinois at Urbana-Champaign, Urbana, Illinois 61801, U.S.A.}
\affiliation[b]{Department of Physics, Kenyon College, Gambier, Ohio 43022, U.S.A.}
\affiliation[c]{Department of Physics, Case Western Reserve University, Cleveland, Ohio 44106, U.S.A.}

\emailAdd{adshead@illinois.edu}
\emailAdd{giblinj@kenyon.edu}
\emailAdd{tscully2@illinois.edu}
\emailAdd{esfaki@illinois.edu}

\abstract{We study the onset of the reheating epoch at the end of axion-driven inflation where the axion is coupled to an Abelian, $U(1)$, gauge field via a Chern-Simons interaction term. We focus primarily on $m^2\phi^2$ inflation and explore the possibility that preheating can occur for a range of coupling values consistent with recent observations and bounds on the overproduction of primordial black holes. We find that for a wide range of parameters preheating is efficient. In certain cases the inflaton transfers all of its energy to the gauge fields within a few oscillations. In most cases, we find that the gauge fields on sub-horizon scales end preheating in an unpolarized state due to the existence of strong rescattering between the inflaton and gauge-field modes. We also present a preliminary study of an axion monodromy model coupled to $U(1)$ gauge fields, seeing a similarly efficient preheating behavior as well as indications that the coupling strength has an effect on the creation of oscillons. % \\
}

\begin{document}
\maketitle
\flushbottom

\section{Introduction}
\label{sec:intro}

Inflationary model building has entered a particularly exciting phase with the demonstration by the BICEP2 experiment \cite{Ade:2014xna} of the sensitivity to B-mode polarization of the cosmic microwave background (CMB) at a level where interesting constraints can be, or soon will be, placed on the inflationary model space. The recent Planck results on dust emission \cite{Adam:2014bub} combined with the joint Planck and BICEP2/KECK Array analysis \cite{Ade:2015tva} mean that dust-foregrounds need to be accurately characterized in order to determine if any of the B-mode polarization observed by the BICEP2 experiment is due to primordial gravitational waves. If confirmed, the observation of these gravitational waves present a spectacular confirmation of one of the early observational predictions for slow-roll inflation \cite{Mukhanov:1981xt}. However, they also present challenges for inflationary model building. The large primordial gravitational wave amplitudes required to explain the BICEP2 signal generically require that the scalar-inflaton field rolls over a distance in field space that is large compared to the four-dimensional Planck scale \cite{Lyth:1996im}. This makes the theory extremely sensitive to unknown physics at short wavelengths, e.g. in the ultra-violet (UV), at or near the Planck scale, and leads to a loss of predictive power, if not a loss of the inflationary mechanism itself. A possible way around this UV sensitivity problem is to find a symmetry powerful enough to forbid interactions between the sector driving the inflationary expansion and other unknown physics.

A promising candidate for the inflaton field is a pseudo-scalar or axion. These fields enjoy shift-symmetries that protect their role as inflatons from being spoiled by coupling to unknown UV physics. Shift-symmetries require that the theory is invariant under a constant shift of the field value, and severely restrict the form of possible interactions with other fields. One of the earliest proposed axion inflation models was Natural inflation \cite{Freese:1990rb, Adams:1992bn}. In this scenario, a cosine potential for the axion is generated by the condensation of a non-Abelian gauge group. Slow-roll inflation is achieved by the hierarchy between the height and width of the potential, the form of which is protected by the shift-symmetry. In order to generate density fluctuations with a spectrum that reproduces the observed fluctuations in the CMB, the axion is required to have a periodicity, or associated mass scale, larger than the Planck scale. This super-Planckian periodicity makes it extremely difficult to embed Natural inflation in a fundamental theory such as string theory \cite{Banks:2003sx}. Model builders sought to circumvent this obstruction in a model dubbed $N$-flation \cite{Dimopoulos:2005ac, Easther:2005zr, Bachlechner:2014hsa} by using a large number of axions, each with small periodicities, to implement assisted inflation \cite{Liddle:1998jc}. While each of the $N$ axions only rolls a small distance in field space, their collective motion is responsible for inflation and effectively traverses a large distance. Unfortunately the $N$ axions in $N$-flation have the effect of renormalizing the Planck scale to a lower value, and the theory ultimately suffers from similar pathologies as the original Natural inflation model it was supposed to fix. Other formulations making use of misaligned axions have also been proposed \cite{Kim:2004rp, Long:2014dta, Burgess:2014oma}. More recently, the observation that a single axion undergoing monodromy can have a small periodicity while ultimately traversing a large field range \cite{McAllister:2008hb, Silverstein:2008sg} has seen resurgent interest in axion inflation \cite{McAllister:2014mpa, Kaloper:2011jz, Marchesano:2014mla, Blumenhagen:2014gta, Hebecker:2014eua,  Cai:2014vua}. For a recent review of axion inflation see Ref.\ \cite{Pajer:2013fsa} and its realizations in string theory see Ref.\ \cite{Baumann:2014nda}.

At the end of the inflationary phase, the Universe must undergo a phase transition, from its super-cooled state to a state filled with radiation and ultra-relativistic matter, to begin the hot big bang. The physics of this phase transition is thought to be highly non-linear, and its details are unknown (see e.g.\ Ref.\ \cite{Amin:2014eta} for a recent review). The shift-symmetry in axion-driven inflation that is so effective at protecting the form of the inflationary sector from unknown UV physics also severely constrains the form of its couplings to the visible sector. These couplings to the standard model of particle physics, either directly or indirectly via intermediaries, are required in order to transfer the inflaton energy into radiation and ultra-relativistic matter.  The shift symmetry dictates that couplings to matter fields must be derivative interactions, therefore a class of allowed interactions are those in which the axion is coupled to a gauge field through a dimension-five Chern-Simons term or Pontryagin density. A coupling of this form is allowed by the symmetries and, from the viewpoint of an effective field theory, must be present. Furthermore, this coupling provides a perturbative decay channel for the axion into gauge bosons which guarantees that reheating eventually completes through perturbative decays alone, and thus provides a viable pathway through which the Universe can transition from inflation into the hot big bang.

The effects of the coupling of axions to gauge fields during inflation is, by now, a well studied field.  It has been known for a long time that axion-gauge field couplings lead to parity-violating gauge-fields that are amplified during slow-roll inflation \cite{Carroll:1991zs, Garretson:1992vt}. The behavior of these gauge fields during inflation and their influence on the inflationary dynamics has also been extensively studied \cite{Prokopec:2001nc, Anber:2009ua, Barnaby:2011vw, Barnaby:2011qe,Ferreira:2014zia, Adshead:2013qp}. Further, the authors of Refs.\ \cite{Shiraishi:2013kxa, Cook:2013xea} studied metric fluctuations generated by a rolling auxiliary pseudo-scalar during inflation. Some work has also been done on the reheating of axion-driven inflation. The authors of Refs.\ \cite{Brandenberger:2008kn} considered stringy reheating of a monodromy scenario while Ref.\ \cite{Blumenhagen:2014gta} studied perturbative decay of an axion inflaton to photons/gauge bosons. Furthermore, previous studies have considered perturbative analyses of the Mathieu equation for $U(1)$ gauge fields coupled to an oscillating scalar or collection of scalars  \cite{ArmendarizPicon:2007iv, Braden:2010wd}. However, these studies were focussed primarily on ranges of parameters considered natural for scenarios of Natural inflation and $N$-flation and concluded that highly non-linear effects and parametric resonance were unimportant in these models. 

Recently, larger couplings between the axion and gauge sectors have been considered \cite{Anber:2009ua, Barnaby:2011qe}, which can lead to observable effects during inflation due to the rescattering of the gauge fields off the axion condensate \cite{Barnaby:2010vf}. Effects such as non-Gaussianity of the density fluctuations, chiral gravitational waves, and the production of primordial black holes  \cite{Barnaby:2010vf, Barnaby:2011vw, Barnaby:2011qe, Linde:2012bt, Bugaev:2013fya} place upper-bounds on the strength of the couplings, the most stringent being due to restrictions on the production of primordial black holes at the end of inflation. 

Our current work addresses a specific problem; given that high-scale inflation requires highly constrained couplings, can the Universe undergo a preheating phase (either tachyonic or resonant) following axion inflation?  Numerical investigations of reheating in canonical \cite{Traschen:1990sw,Kofman:1994rk,GarciaBellido:1997wm,Khlebnikov:1997di,Greene:1997ge,Parry:1998pn,Bassett:1998wg,GarciaBellido:1998gm,Easther:1999ws,Liddle:1999hq,Finelli:2001db,Bassett:2005xm,Podolsky:2005bw} or non-canonical \cite{Child:2013ria} scalar field scenarios is becoming routine. Furthermore, couplings of scalar fields to Abelian \cite{Rajantie:2000fd,Copeland:2002ku,Deskins:2013dwa} or non-Abelian gauge fields in cosmological settings have been studied in recent years \cite{GarciaBellido:1999sv,Smit:2002yg,Tranberg:2003gi,Skullerud:2003ki,GarciaBellido:2003wd,vanTent:2004rc,vanderMeulen:2005sp,Tranberg:2006dg,Dufaux:2010cf,GarciaBellido:2008ab,DiazGil:2007dy,DiazGil:2008tf, Mazumdar:2008up, Allahverdi:2011aj, Adshead:2012kp, Maleknejad:2012fw}.  The question we address here is new.  We employ lattice simulations, using the same numerical technique of Ref.\ \cite{Deskins:2013dwa}, to study the possibility of tachyonic or parametric amplification of gauge fields after  inflation due to the presence of a Chern-Simons term.  As in most studies of preheating, we remain agnostic as to the specific nature of the coupled gauge field and do not necessarily identify it as a $U(1)$ gauge field from the standard model.

Our results can be summarized as follows. We find that for reasonable ranges of the axion-gauge field coupling, non-linear effects can be very important at the end of inflation. In particular, at the middle to upper range of the couplings allowed by black hole abundance, we find that reheating is essentially instantaneous, proceeding via a phase of tachyonic resonance \cite{Dufaux:2006ee} and completing within a single oscillation of the axion. Despite the asymmetry in the equations of motion for the two polarizations of the gauge fields, on sub-horizon scales, rescattering of the gauge bosons off the axion condensate is efficient at generating the second polarization. On scales larger than the horizon at the end of inflation, an asymmetry between the gauge field polarizations remains. The Universe that results in these cases is radiation dominated and is characterized by a very high reheating temperature.  As the coupling is decreased, this tachyonic resonance is weakened and the axion oscillates multiple times before reheating completes. During these multiple oscillations equal levels of both polarizations of the gauge field are excited. Decreasing the coupling further yields a brief window where parametric resonance effects become important before preheating abruptly shuts off and non-linear effects cease to be important. At these lower couplings, non-linear effects are negligible and the Universe reheats via perturbative decay of the axion into gauge bosons.

We also investigate how these preheating effects might depend on the shape of the inflationary potential.  As a second test, we subject the axion to a monodromy-type potential \cite{McAllister:2008hb}, and show that the range of couplings for which efficient preheating can occur is slightly different, although the same order-of-magnitude.  We conclude by presenting an intriguing set of data that suggest gauge fields might play a role in the creation and evaporation of oscillons in this scenario.

We work in natural units where $\hbar = c =1$, however, we retain the Planck mass, $m_{\rm pl}=1.22 \times 10^{19}\,{\rm GeV}$.

%%%%%%%%%%%%%%%%%%%%%%%%%%%%%%%%%%%%%%%
%%%%%%%%%%%%%%%%%%%%%%%%%%%%%%%%%%%%%%%
%%%%%%%%%%%%%%%%%%%%%%%%%%%%%%%%%%%%%%%

\section{Background and conventions}\label{background}

We begin with the usual action for axion-driven inflation
\begin{align}\label{eqn:axionact}
\mathcal{S}_{\rm inf} = \int d^4 x \sqrt{-g}\[\frac{m_{\rm pl}^2}{16 \pi}R - \frac{1}{2}\partial_\mu\phi\partial^\mu \phi - V(\phi)\],
\end{align}
where $\phi$ is a pseudo-scalar (axion) and $V(\phi)$ is a potential that supports slow-roll inflation.  For definiteness, we consider the potentials for the simplest type of chaotic inflation \cite{Linde:1981mu},
\begin{align}
\label{eqn:quadpot}
V(\phi) = & \frac{1}{2}m^2\phi^2,
\end{align}
and the simplest type of axion monodromy inflation \cite{McAllister:2008hb}
\begin{align}\label{eqn:monopot}
V(\phi) = & \mu^3 \(\sqrt{\phi^2+\phi_c^2} - \phi_c\),
\end{align}
which is well described by a linear function of $\phi$ for large field values.  The amplitude of the scalar spectrum fixes the parameters $m$ and $\mu$ to be \cite{Ade:2013uln, Peiris:2013opa, Easther:2013kla}
\begin{equation}
m \approx  1.06 \times 10^{-6}\,m_{\rm pl},
\end{equation}
and
\begin{equation}
\mu \approx  1.20 \times 10^{-4} \,m_{\rm pl}.
\end{equation} 
The parameter $\phi_c$ in the monodromy potential, Eq.~\ref{eqn:monopot}, has a negligible effect on the spectrum of curvature fluctuations and gravitational waves on the scales that are observable in the CMB (provided, of course, that $\phi_c$ is much smaller than the field value where the CMB fluctuations are generated) and, consequently, is unconstrained by data. However, $\phi_c$ becomes important near the end of inflation and has a small effect on the value of $\phi$ when inflation ends. Further, 
for small field values, $\phi < \phi_c$, the potential can be expanded as
\begin{align}
V(\phi) \approx \frac{\mu^3}{2|\phi_c|}\phi^2 - \frac{\mu^3}{8|\phi_c|^3}\phi^4+\ldots\,,
\end{align}
and  the resulting dynamics of $\phi$ in this region depend strongly on the value of $\phi_c$ \cite{Amin:2011hj}.

In addition to the axion, we consider a $U(1)$ gauge field coupled to the axion
\begin{align}\label{eqn:gaugeact}
\mathcal{S}_{\rm gauge} = & \int d^4 x \sqrt{-g}\[-\frac{1}{4}F_{\mu\nu}F^{\mu\nu} - \frac{\alpha}{4 f}\phi F_{\mu\nu}\tilde{F}^{\mu\nu}\],
\end{align}
where $\alpha$ is a dimensionless coupling constant of order unity and $f$ is a mass scale associated with the pseudo-scalar (axion). The field strength and its dual are given by the standard expressions, $F_{\mu\nu} = \partial_\mu A_\nu - \partial_{\nu}A_{\mu}$ and $\tilde{F}^{\mu\nu} = \epsilon^{\mu\nu\alpha\beta}F_{\alpha\beta}/2$,
where $\epsilon^{\mu\nu\alpha\beta}$ is the completely antisymmetric tensor, and our convention is
\begin{align}
\epsilon^{0123} = \frac{1}{\sqrt{-g}}.
\end{align}
Greek letters here and throughout denote four dimensional spacetime indices and Roman letters from the middle of the alphabet are used to denote spatial indices. Repeated lower spatial indices are summed using the Kronecker delta. We work with the Friedmann-Lema\^itre-Robertson-Walker (FLRW) metric in conformal time with mostly-plus conventions
\begin{align}
ds^2 = -a^2(d\tau^2 - d {\bf x}^2).
\end{align}
The equation of motion for the pseudo-scalar field is the Klein-Gordon equation sourced by the Chern-Simons density of the gauge field
\begin{align}
\label{axioneom}
(\partial_\tau^2 + 2\mathcal{H}\partial_\tau - \partial_i \partial_i)\phi + a^2 \frac{dV}{d\phi} = \frac{\alpha}{4 f}a^2 F_{\mu\nu}\tilde{F}^{\mu\nu},
\end{align}
where $\mathcal{H} = a'/a$ and where here and below $' \equiv \partial_\tau = \partial/\partial \tau$. The equations of motion for the gauge field are 
\begin{align}
\partial_{\rho}\(\sqrt{-g}F^{\rho\sigma}\)+\frac{\alpha}{f} \partial_{\rho}(\sqrt{-g}\phi \tilde F^{\rho\sigma}) = 0.
\end{align}
The $\sigma = 0$ equation is the Gauss' law constraint
\begin{align}
\label{gaugeeom1}
\partial_j \partial_j A_0  - \partial_\tau  \partial_{i} A_i+\frac{\alpha}{f}\epsilon_{ijk} \partial_{k}\phi \partial_i A_j  = 0,
\end{align}
while the $\sigma = i$ equations are the field equations for the spatial components of the gauge field
\begin{align}%\nn
\label{gaugeeom2}
-\partial_{\tau}\(  \partial_\tau A_i - \partial_i A_0 \)+\partial_{m} (\partial_m A_i - \partial_i A_m)+\frac{\alpha}{f} \epsilon_{imk}\partial_{\tau}\phi \partial_m A_k  %& \\ 
-\frac{\alpha}{f} \epsilon_{ i m k}\partial_{m}\phi (\partial_\tau A_k - \partial_k A_0) =  & 0.
\end{align}
Finally, assuming the metric is unperturbed, the scale factor satisfies Einstein's equations
\begin{equation}
\label{ffriedman}
\frac{3 m_{\rm pl}^{2} }{8\pi}\mathcal{H}^2 =  a^2\rho, 
\end{equation}
and
\begin{equation} % \\
\frac{m_{\rm pl}^{2}}{8\pi}\(\mathcal{H}' - \mathcal{H}^2\) =  -a^2\frac{\rho+p}{2}.
\end{equation}
The pressure, $p$, and energy  density, $\rho$, are found from the stress-energy tensor
\begin{align}
 T_{\mu\nu} = &\tr\[F_{\mu\alpha}F_{\nu\beta}\]g^{\alpha\beta}  -\frac{g_{\mu\nu}}{4}F_{\mu\nu}F^{\mu\nu}  -g_{\mu\nu}\left[\frac{1}{2}g^{\rho\sigma}\partial_{\rho}\phi\partial_{\sigma}\phi+V(\phi)\right] + \partial_{\mu}\phi\partial_{\nu}\phi,
\end{align}
which can be explicitly written as
\begin{align}
\rho = & \frac{1}{2}\frac{\phi'{}^2}{a^2} +\frac{1}{2}\frac{(\partial_i \phi)^2 }{a^2}+  V(\phi) + \frac{1}{2 a^4}(\partial_0 A_i - \partial_i A_0)^2 + \frac{1}{4 a^4}(\partial_i A_j - \partial_j A_i)^2,
\end{align}
and
\begin{align}
p  = &  \frac{1}{2}\frac{\phi'{}^2}{a^2} +\frac{1}{2}\frac{(\partial_i \phi)^2 }{a^2}-  V(\phi) + \frac{1}{6 a^4}(\partial_0 A_i - \partial_i A_0)^2 + \frac{1}{12 a^4}(\partial_i A_j - \partial_j A_i)^2.
\end{align}
Note that the axion-gauge field coupling does not contribute directly to the stress-energy tensor.

%%%%%%%%%%%%%%%%%%%%%%%%%%%%%%%%%%%%%%%
%%%%%%%%%%%%%%%%%%%%%%%%%%%%%%%%%%%%%%%
%%%%%%%%%%%%%%%%%%%%%%%%%%%%%%%%%%%%%%%

\section{Gauge-field production during inflation}\label{sec:duringinflation}

During inflation, the coupling of the gauge field to the axion results in exponential production of one polarization of the gauge field over the other \cite{Carroll:1991zs, Garretson:1992vt, Prokopec:2001nc}. To see this, we first fix  the gauge  by choosing Coulomb, or transverse, gauge $\partial_i A_i = 0$. The Gauss' law constraint, Eq.~\ref{gaugeeom1},  then implies that $A_0 = 0$ at linear order in fluctuations.\footnote{Note that at this order (linear) in fluctuations, Coulomb gauge and temporal gauge ($A_0 = 0$) are equivalent. This can be seen trivially from Eqn.\ \eqref{gaugeeom1}, the Gauss' law constraint. In the linear regime, this equation reads $\partial_j\partial_j A_0 - \partial_\tau \partial_j A_j = 0$. In Coulomb gauge, assuming $k \neq 0$, this constraint reads $A_0 = 0$. In temporal gauge, Gauss's law reads $\partial_\tau\partial_j A_j = 0$, which for $k\neq 0$, implies $\partial_j A_j = 0$.}  

At linear order in fluctuations, with this choice of gauge, the equation of motion for the gauge field becomes
\begin{align}\label{eqn:gauge field}
\partial^2_{\tau}A_i - \partial_m\partial_m A_i - \frac{\alpha}{f}\epsilon_{imk}\partial_\tau \phi \partial_m A_k = 0 .
\end{align}
To study the fluctuations of the gauge field, it proves most convenient to work in Fourier space, where our convention is
\begin{align}
\vec{A}({\bf x})  = & \int \frac{d^3 k}{(2\pi)^3} \vec{A}_{\bf k}  e^{i {\bf k}\cdot \bf{x}}.
\end{align}
The Fourier components are then expanded in a basis of helicity states 
\begin{align}
\vec{A}_{\bf k} = \sum_{\lambda = \pm}A^{\lambda}_{\bf k} \vec{\varepsilon}
\,{}^{\lambda}({\bf k }),
\end{align}
where the polarization vectors $ \varepsilon^{\lambda}_i({\bf k })$ satisfy the orthogonality and normalization relations
\begin{align}\nn
k_i \varepsilon^{\pm}_{i}({\bf k}) = & 0,\quad
\epsilon^{ijk}k_{j} \varepsilon^{\pm}_{k}({\bf k})  =  \mp i k \varepsilon^{\pm}_{i}({\bf k}), \\
\varepsilon_i^\pm({\bf k})^* = & \varepsilon_i^{\pm}(-{\bf k}), \quad 
 \varepsilon^{\lambda}_{i}({\bf k})\varepsilon^{\lambda'}_{i}(-{\bf k})  =  \delta_{\lambda \lambda'}.
\end{align}
In this situation, conformal time is defined to be a negative, increasing quantity during inflation
\begin{align}
d\tau = \frac{dt}{a}, \quad \tau  = & \int_t \frac{dt}{a} = \int \frac{d\ln a}{a H}\approx - \frac{1}{aH},
\end{align}
where the last approximation is exact in the de Sitter limit, $\epsilon_H \to 0$, where the slow-roll parameter, $\epsilon_H$ is defined as $\epsilon_H = -\dot{H}/H^2$. Here and throughout, an overdot is used to denote a derivative with respect to cosmic time, $t$. 

We can now quantize the modes by introducing the creation and annihilation operators, $a_{\lambda}({\bf k})$ and $a_{\lambda}^{\dagger}({\bf k})$ satisfying the canonical commutation relations
\begin{align}
\[a_{\lambda}({\bf k}), a^{\dagger}_{\lambda'}({\bf k'})\] = (2\pi)^{3}\delta_{\lambda\lambda'}\delta^{3}({\bf k} - {\bf k}'),
\end{align}
which allows us to expand the mode-functions as
\begin{align}
A_i(\tau, {\bf x}) = & \sum_{\lambda = \pm}\int \frac{d^3 k}{(2\pi)^3}e^{i {\bf k}\cdot {\bf x}}\varepsilon^{\lambda}_{i}({\bf k}) \[A^{\lambda}(k, \tau) a_{\lambda}({\bf k}) +A^{\lambda,*}(k, \tau)a^{\dagger}_{\lambda}(-{\bf k}) \].
\end{align}
With our conventions, the gauge field equation of motion Eq.~\ref{gaugeeom2} becomes a separate equation for each polarization, depending only on the magnitude of the momenta $k = |{\bf k}|$
\begin{align}\label{eqn:kspacegfeqn}
\(\partial_{\tau}^2 + k^2 \pm \frac{\alpha}{f}\frac{\dot{\phi}}{H} \frac{k}{\tau}\) A^{\pm}_{k}= 0,
\end{align}
where we have also made use of the de Sitter approximation for the scale factor during inflation in the last term. During inflation, $\epsilon_{H} = 4\pi \dot{\phi}^2/( H^2 m_{\rm pl}^2)\approx {\rm const.}$ and, after changing variable to $u = 2 i k\tau $, the equation of motion is transformed to the Whittaker equation
\begin{align}
\(\frac{d^2}{dz^2} - \frac{1}{4}+\frac{\lambda}{z}+\frac{1/4 - \mu^2}{z^2}\)W_{\lambda, \mu}(z) = 0.
\end{align}
In our case, we have
\begin{align}
\(\partial_u^2 - \frac{1}{4}\mp i \frac{\gfc}{u}\)A_{k}^{\pm} = 0,
\end{align}
and thus $\mu = 1/2$ and $\lambda = \mp i \gfc$ where we have defined
\begin{align}\label{eqn:xidef}
\gfc =  \frac{1}{2}\frac{\alpha}{f}\frac{\dot{\phi}}{H} %={\rm sign}(\dot\phi) M_{\rm pl} \frac{\alpha}{f}\sqrt{\frac{\epsilon_H}{2}} 
= {\rm sign}(\dot\phi) {m_{\rm pl}\over \sqrt{8\pi}} \frac{\alpha}{f}\sqrt{\frac{\epsilon_H}{2}} .
\end{align}
The general solution of the Whittaker equation can be written in terms of the Whittaker W-function
\begin{align}
A_{k}^{\pm}(k\tau) = C_1 W_{\mp i \gfc, \frac{1}{2}}(2 i k \tau) + C_2 W_{\pm i \gfc, \frac{1}{2}}(-2 i k \tau) .
\end{align}
The constants of integration, $C_1$ and $C_2$, are set by canonical quantization which amounts to normalizing the modes according to the Wronskian condition
\begin{align}
W[A^{\pm}_k(k\tau), (A^{\pm}_{k}(k\tau))^*] = i,
\end{align}
and demanding that the modefunction approaches the Minkowski vacuum in the limit, $k|\tau| \to \infty$
\begin{align}
\lim_{k|\tau| \to \infty} A^{\pm}({k}, \tau) =  \frac{1}{\sqrt{2k}}e^{-i k \tau \mp i \gfc \ln (-2 k \tau)}.
\end{align}
The properly normalized solutions are 
\begin{align}
A^{\pm}(k, \tau) = \frac{e^{\pm \frac{\pi}{2}\gfc}}{\sqrt{2 k}}W_{ \mp i\gfc, \frac{1}{2}}(2 i k\tau).
\end{align}
In the limit $k|\tau| \to 0$, the asymptotic form of the mode function is
\begin{align}
\lim_{k|\tau| \to 0} A^{\pm}(k, \tau) =  \frac{1}{\sqrt{2k}}\frac{e^{\pm \frac{\pi}{2}\gfc}}{\Gamma(1\pm i \gfc)}.
\end{align}
Compared to the conformally invariant radiation solution, the circularly polarized modes get amplified by a factor
\begin{align}
\left| \frac{A^{\pm}}{A^{\pm,\rm rad}}\right| \simeq e^{\frac{\pi}{2}|\gfc|\pm \frac{\pi}{2}\gfc},
\end{align}
where we have used the Stirling formula
\begin{align}
|\Gamma(1\mp i\gfc)| \simeq (2\pi |\gfc|)^{1/2}e^{-\pi |\gfc|/2},
\end{align}
and we have assumed that $|\gfc| > 1$. Note that this means that when $\gfc > 0$ ($\gfc < 0$), the mode $A^{+}_k$ ($A^{-}_k$) gets amplified by a factor $\sim e^{\pi |\gfc|}$ while the other mode is unchanged. For this work, we focus on large-field inflationary models, and assume that $\dot\phi < 0 $ so that the mode $A^-$ is amplified during inflation.

The exponentially enhanced gauge fields have important effects during inflation due to their re-scattering off the inflaton condensate and their interactions with the metric. The former leads to the production of fluctuations of the inflaton which are statistically non-Gaussian, while the latter leads to the production of gravitational radiation \cite{Barnaby:2011qe}. The Gaussianity of the observed density fluctuations by Planck \cite{Ade:2013ydc} then implies that the quantity $\gfc_{\rm CMB} \lesssim 2.22$ \cite{Pajer:2013fsa}, where $\gfc_{\rm CMB}$ is the quantity in Eq.~\ref{eqn:xidef} evaluated during the time when the modes that form the CMB leave the horizon.

During inflation, $\epsilon_H$ and thus the ratio $\dot\phi/(H m_{\rm pl})$ increases. This means that shorter-wavelength modes that leave the horizon later in inflation are amplified more than their longer-wavelength counterparts that leave the horizon earlier. The largest effects occur when $\epsilon_H$ is near unity near the end of inflation.  In the limit that $\gfc \gg 1$ (which for models that satisfy $\gfc_{\rm CMB} < 2.22$ only possibly occurs near the end of inflation) the energy density in the gauge fields becomes important and the gauge-field fluctuations begin to backreact on the homogeneous background equations of motion. In this limit, in the Hartree approximation, the Friedmann (Eq.~\ref{ffriedman}) and Klein-Gordon equations (Eq.~\ref{axioneom}) become
\begin{align}\label{eqn:backreactfried}
\frac{3 m_{\rm pl}^{2} }{8\pi}\mathcal{H}^2 =  \frac{\phi'{}^2}{2}+a^2V(\phi) + \frac{a^2}{2}\langle E^2+B^2\rangle,
\end{align}
\begin{align}
\frac{m_{\rm pl}^{2}}{8\pi}\(\mathcal{H}' - \mathcal{H}^2\) =  -\(\frac{\phi'{}^2}{2} +\frac{2}{3}a^2\langle E^2+B^2\rangle\),
\label{eqn:friednmaneq}
\end{align}
and
\begin{align}
\label{eqn:backreactKG}
{\phi}''+2\mathcal{H}\phi' + a^2V' =  \frac{\alpha}{f} a^2 \langle {\bf E} \cdot {\bf B} \rangle,
\end{align}
where the electric and magnetic fields\footnote{We refer to these fields as electric and magnetic, however, they need not be the electro-magnetic fields of the standard model.} associated with the $U(1)$ gauge field are $E_i = a^{-2}A'_i$ and $B_i = a^{-2} \epsilon_{ijk}\partial_j A_k$. In this limit, up to an irrelevant constant phase, the gauge field mode that is amplified is approximated near horizon crossing by \cite{Anber:2009ua}
\begin{align}
A^{-}_k (\tau) = \frac{1}{\sqrt{2k}}\(\frac{k|\tau|}{2|\gfc|}\)^{1/4}\exp\(\pi|\gfc| - 2\sqrt{2|\gfc| k|\tau|} \),
\end{align}
while the other mode is unaffected and is negligible. The expectation values of the quantum fields are well approximated by \cite{Anber:2009ua}\footnote{Note that the axion velocity assumed here is opposite to that assumed by \cite{Anber:2009ua}. This means relative to this work, the  other gauge mode is amplified and consequently  the sign of $\langle {\bf E} \cdot {\bf B} \rangle$ is opposite.}
\begin{align}\label{eqn:GFenergy}
\frac{1}{2}\langle E^2+B^2\rangle \simeq & 1.4 \cdot 10^{-4}\frac{H^4}{|\gfc|^3}e^{2\pi| \gfc|}, \quad
\langle {\bf E} \cdot {\bf B} \rangle \simeq  2.4 \cdot 10^{-4} \frac{H^4}{|\gfc|^4}e^{2\pi|\gfc|}.
\end{align}
Toward the end of inflation, for large values of $m_{\rm pl}\, \alpha/f$, the back reaction of the gauge fields on the rolling axion becomes important and inflation is prolonged \cite{Barnaby:2011qe}. During this phase, the primordial density fluctuation spectrum is expected to be dominated by rescattering and large, non-Gaussian density fluctuations are predicted. These large amplitude density fluctuations can produce primordial black-holes \cite{Carr:2009jm, Lin:2012gs, Alabidi:2012ex} and ensuring that they are not overproduced requires that $\gfc_{\rm CMB} \lesssim  1.5 - 1.7$ for $m^2 \phi^2$ \cite{Linde:2012bt, Bugaev:2013fya} and $\gfc_{\rm CMB} \lesssim  1.8$ for monodromy  \cite{Bugaev:2013fya}, which is tighter than the current bounds from the Gaussianity of the CMB fluctuations. These limits are model dependent, and are somewhat sensitive to the form of the potential. In this work we consider only values of the coupling $\alpha/f$ such that $\gfc_{\rm CMB} < 1.5 - 1.7$. This bound translates to roughly $\alpha/f \lesssim 110\,m_{\rm pl}^{-1} - 125\,m_{\rm pl}^{-1}$ for the $m^2 \phi^2$ potential and $\alpha/f \lesssim 180\,m_{\rm pl}^{-1}$ for the simple monodromy potential of Eq.~\ref{eqn:monopot}. In practice, we do not approach this threshold in our simulations, keeping the couplings used for our simulations lower by around a factor of two.

The bounds from primordial black-hole abundances rely on the above approximations of Eqs.~\ref{eqn:backreactfried} - \ref{eqn:backreactKG} together with Eq.~\ref{eqn:GFenergy}   being a good description of the system in the region of strong back reaction \cite{Linde:2012bt}. However, these approximations do not yield a self consistent set of equations because the resulting stress-energy tensor is not covariantly conserved. This means that in the region where back reaction becomes significant, the above equations are inaccurate.  For small couplings, we expect that the above approximations are accurate enough to capture the onset of back reaction. In what follows we initialize our lattice simulations using the values of the fields   found from the numerical evolutions of the Eqs.~\ref{eqn:backreactfried} - \ref{eqn:backreactKG}. We make use of the approximations of Eq.~\ref{eqn:GFenergy}  for the expectation values of the energy density in gauge field fluctuations and the Pontryagin density respectively. The use of these approximations implies that the initialization of our simulations is less and less accurate for larger values of the coupling between the gauge field and the axion. To minimize this error, we initialize our simulations two e-foldings before inflation ends. The modes that are important in the reheating era are well within the horizon at this time, and well described by the linear approximation at the start of our simulations (See Table \ref{tab:phi0} and Fig.~\ref{fig:initmodes}).

%%%%%%%%%%%%%%%%%%%%%%%%%%%%%%%%%%%%%%%
%%%%%%%%%%%%%%%%%%%%%%%%%%%%%%%%%%%%%%%
%%%%%%%%%%%%%%%%%%%%%%%%%%%%%%%%%%%%%%%

\section{Perturbative reheating}
\label{sec:perturbative}

Before we move to the non-linear regime, we briefly revisit the perturbative reheating case. The coupling of the axion to gauge fields provides a natural decay channel for the axion to produce gauge bosons. Even in the absence of non-linear effects due to the homogenous motion of the inflaton condensate, the Universe will reheat to gauge bosons  via perturbative decays of the inflaton into a pair of gauge bosons. When the Hubble rate becomes comparable to the perturbative decay rate, the decay of the inflaton into gauge bosons proceeds rapidly. The contribution of these newly created particles to the energy density quickly dominates and reheats the universe. This  represents a lower bound on the temperature of the Universe at reheating.

The $\phi F\tilde F$ coupling allows the axion to decay to two gauge bosons with a rate that is well known (see e.g.\ \cite{Agashe:2014kda})
\begin{align}
\Gamma_{\phi \to A A} = \frac{\alpha^2 m_{\phi}^3}{64\pi f^2},
\end{align}
where $m_{\phi}$ is the mass of the axion about its vacuum. This rate sets a lower bound on the reheating temperature which is found by comparing the decay rate to the Hubble rate; perturbative reheating finishes when $\Gamma/3H \sim 1$,  which results in a reheating temperature
\begin{align}
T_{\rm reh} \sim & \(\frac{5}{4\pi^3 g_*}\)^{1/4}\sqrt{\Gamma m_{\rm pl}}
\simeq  0.14 \(\frac{100}{g_*}\)^{1/4}\sqrt{\Gamma m_{\rm pl}},
\end{align}
where $g_*$ is the number of relativistic degrees of freedom. For the simplest version of chaotic inflation, we can write 
\begin{align}
T_{\rm reh}\sim  1.31 \times 10^{9} \(\frac{100}{g_*}\)^{1/4}\(\frac{m_{\phi}}{1.06 \times 10^{-6}m_{\rm pl}}\)^{3/2}\(\frac{\alpha/(f m_{\rm pl})}{10}\) \quad {\rm  GeV}.
\end{align}
For small values of the coupling, this is the dominant effect. Reheating in this case proceeds by perturbative decay of the axion into gauge bosons.

%%%%%%%%%%%%%%%%%%%%%%%%%%%%%%%%%%%%%%%
%%%%%%%%%%%%%%%%%%%%%%%%%%%%%%%%%%%%%%%
%%%%%%%%%%%%%%%%%%%%%%%%%%%%%%%%%%%%%%%

\section{Instabilities and resonance}\label{sec:instabandres}

We are interested in what happens immediately following the inflationary epoch. We assume that the pseudo-scalar begins to oscillate about the minimum of its potential, which we initially take to be a quadratic function, Eq.~\ref{eqn:quadpot}.  While this form is exact for the simplest model of inflation, there are anharmonic corrections to the potential for other important axion inflation scenarios, such as monodromy inflation. 

It is instructive to first consider the behavior of the system by temporarily neglecting the expansion of space and considering only the linear theory to gain intuition for the system and to identify features present in the fully nonlinear treatment. In this regime the axion satisfies the equation
\begin{align}
\ddot{\phi} +m^2 \phi = 0,
\end{align}
where overdots denote derivatives with respect to cosmic time $t$. This equation has the simple solution
\begin{align}\label{eqn:phial}
\phi(t) = \phi_0 \cos(m t),
\end{align}
where $\phi_0$ is (approximately) the field value at which the slow-roll conditions are violated and inflation ends. To estimate the effect of parametric resonance (while still neglecting the expansion of space) we write the  equation of motion for the mode amplitudes, Eq.~\ref{eqn:kspacegfeqn}, as
\begin{align}
\label{akeom}
\ddot{A}_k^{\pm} + k\(k  \mp \frac{\alpha}{f} \dot{\phi} \) A^{\pm}_k = 0.
\end{align}
With the solution for $\phi(t)$ from Eq.\ \eqref{eqn:phial}, and after redefining time $z = mt/2$, this equation can be recast as
\begin{align} \label{eqn:gfmat}
\[\frac{d^2 }{dz^2} + 4{k\over m}\( {k\over m} \mp \frac{\alpha}{ f }\phi_0 \cos( 2z)\)\]A^{\pm}_k = 0.
\end{align}
The fact that each helicity obeys a different equation is irrelevant here because the sign of the term proportional to $\alpha/f$ in Eq.~\ref{eqn:gfmat} can be reversed by a constant shift of its argument $z \to z+ \pi/2$. In this approximation both modes are expected to grow equally after each inflaton oscillation. We  show how this symmetry between the two helicities is broken once the expansion of the Universe is taken into account.

We can compare Eq.~\ref{eqn:gfmat} to the normal Mathieu equation,
\begin{equation} \label{eqn:mat}
\frac{d^2 u}{dz^2} + \left[A_k + 2q \cos(2z)\right]u = 0,
\end{equation}
to see that\footnote{This definition of the two parameters of the Mathieu equation as $A_k$ and $q$ is typically used in models where the inflaton decays to scalars. In these models $q$ does not depend on the wavenumber. In the case of gauge fields $q$ does depend on $k$, but we refrain from using a subscript to be consistent with prior literature.}
\begin{equation}
A_k = 4\( {k\over m } \) ^2, \quad
q = \mp 2  {k\over m} \frac{\alpha}{f} \phi_0.
\label{eq:matparams}
\end{equation}

\begin{figure}[h!] 
\centering
\includegraphics[height=0.35\columnwidth]{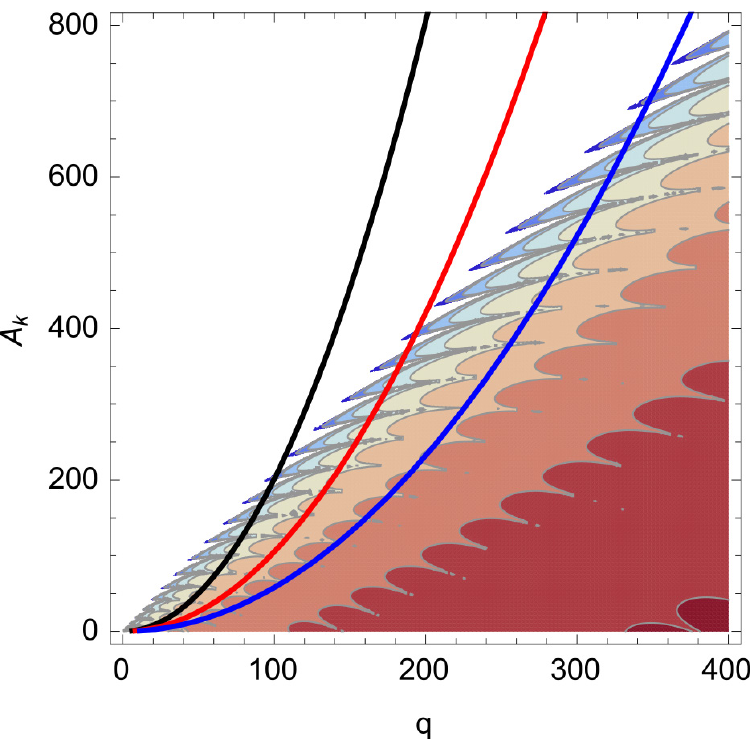} \phantom{sp}
\includegraphics[height=0.35\columnwidth]{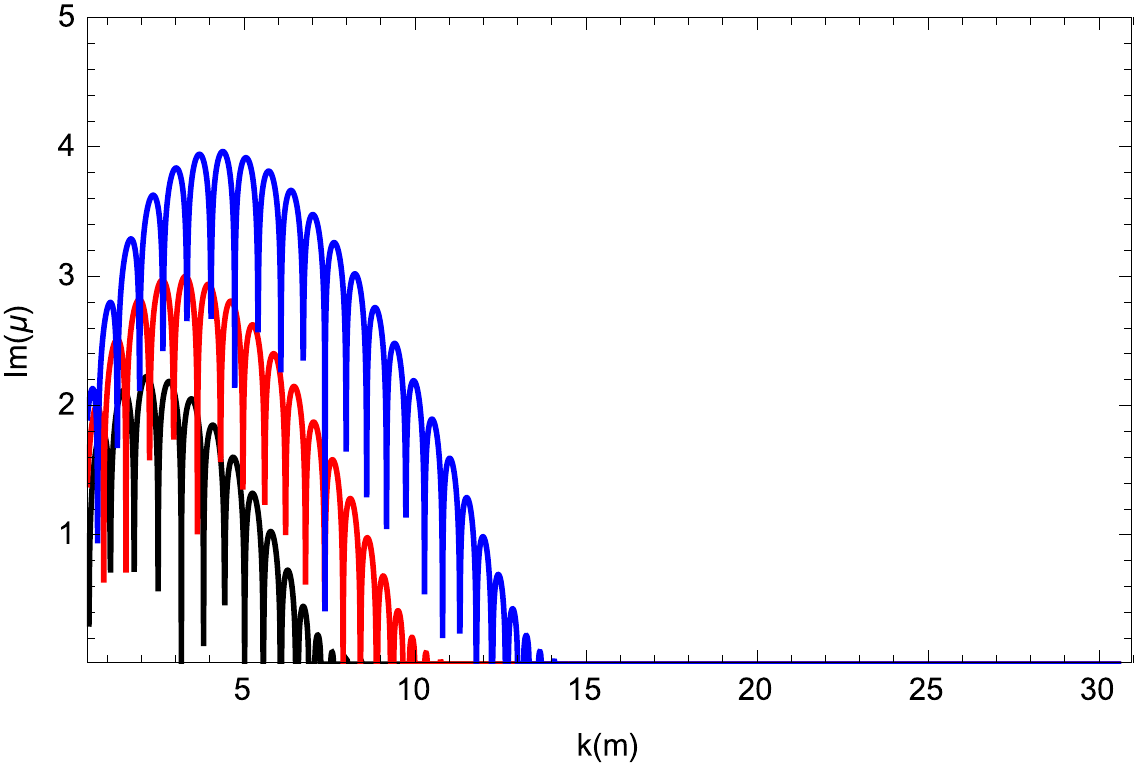}
\caption{A Mathieu instability diagram for the process in question.  The left panel is a contour plot of the magnitude of the imaginary part of the Mathieu characteristic exponent, ${\rm Im}(\mu)$, which paramaterizes the rate of growth of the instability--non-zero values of this quantity indicate modes that grow exponentially.  The three curves on the plot show the slice of parameter space along which the momentum modes in our system lie for three different couplings, $\alpha/f = 35\, m_{\rm pl}^{-1}$ where $\phi_0 \approx 0.20 \,m_{\rm pl}$ (black), $\alpha/f = 45\, m_{\rm pl}^{-1}$ where $\phi_0 \approx 0.22\,m_{\rm pl}$ (red), and $\alpha/f = 55\, m_{\rm pl}^{-1}$ where $\phi_0\approx0.24 \, m_{\rm pl}$ (blue).  The right panel shows the size of the characteristic exponent as a function of the momentum of the mode $k/m$ for the same three couplings.  These curves  identify the modes that are excited during the first oscillation of the field; as the amplitude of the oscillation decreases, these curves move to smaller $q$ and become steeper. The range on the right panel is chosen to correspond to the values of the comoving wavenumbers that can be probed by our simulations.}
\label{fig:matinstab}
\end{figure}

From Eq.~\ref{eqn:mat} we see two important thresholds that define the behavior of the solutions.  These thresholds are the tachyonic resonance threshold, which is set by $A_k < 2q$, and the broad-to-narrow resonance threshold, which is determined by the size of $q$. Broad resonance occurs for $q \gg 1$ while narrow resonance occurs for $q \lesssim1$.  Fig.~\ref{fig:matinstab} shows the Mathieu instability diagram for our process, along with three $\{q,A_k\}$ curves for fixed coupling, $\alpha/f \phi_0$, and varying wavenumber. The $A_k=2q$ line is the diagonal for our choice of axes range. It can be clearly seen that in the regime where $A_k<2q$, the instability bands are much broader and the imaginary parts of the Mathieu exponents are larger. It is also interesting to note that this system can be cast in terms of only two (dimensionless) combinations: the ratio of wavenumber to mass scale, $k/m$, and the product of the coupling strength and initial axion amplitude, $(\alpha/f) \phi_0$.  The curves in Fig.~\ref{fig:matinstab} are defined by a fixed coupling and initial field amplitude $( \alpha/f) \phi_0$, and are parameterized by wavenumber.

When the combination $A_k + 2q\,\cos(2z) < 0$, which occurs for $A_k < 2q$, there is a {\sl tachyonic} instability in the equation of motion \cite{Dufaux:2006ee}.  This condition defines a set of modes
\begin{equation}
\frac{k}{m}<\frac{\alpha}{ f} \phi_0 ,
\end{equation}
whose mass-squared is negative ($u^{\prime\prime} \propto u$ from Eq.~\ref{eqn:mat}).  There is always a set of such modes, as long as the homogeneous mode of the axion is oscillating.  

The details of this tachyonic regime are seen directly from the equations of motion for the two polarizations, Eq.~\ref{akeom}.  When the axion {\sl velocity} changes sign, the combination 
\begin{equation}\label{eqn:tachyonicregime}
k \mp \frac{\alpha}{f} \dot\phi < 0,
\end{equation}
is negative for only {\sl one} of the two polarizations, depending on the sign of $\dot\phi$. When the axion velocity is positive (negative), the $A^+$ ($A^-$) mode is tachyonically amplified.  This regime is obviously most important for large couplings and when the amplitude of the field oscillation is large. In these large coupling cases, preheating is extremely efficient and can complete after only one or two oscillations of the homogeneous inflation condensate. These tachyonic instabilities disappear when the homogeneous mode of the inflation breaks down, which occurs due to back scattering of the gauge modes, or self resonance of the axion itself (when the axion potential includes anharmonic terms).  This should happen very quickly for large couplings -- in some cases, only one polarization is amplified by this tachyonic resonance. However, as we show, in these cases rescattering or backscattering effects are extremely efficient and generate equal amounts of the non-tachyonic polarization.  As can be seen from Fig.~\ref{fig:matinstab}, the wavenumbers corresponding to the tachyonic regime (Eq.~\ref{eqn:tachyonicregime}) lead to much larger Mathieu exponents, so they dominate the behavior of the system, hence we concentrate on them.  

Having discussed the behavior of Eq.~\ref{eqn:mat}  for $A_k<2q$, we move to the second threshold which is defined by the size of $q$. If the $\{q,A_k\}$ curve (with constant $\alpha/f \phi_0$) intersects the Mathieu bands for values of $q\gg 1$, we are in the broad resonance regime \cite{Kofman:1997yn,Dufaux:2006ee} characterized by the curve intersecting large instability bands. In the opposite regime, $q \lesssim 1$, the modes that are amplified have frequencies comparable to those of the oscillating inflaton. This is called the narrow resonance regime, since the instability bands of the Mathieu chart have a narrow width. The growth rate of gauge field modes in this regime can be fully analyzed using the methods of parametric resonance based on Floquet theory, as described for example in \cite{Kofman:1997yn} and applied to narrow resonance for gauge fields in \cite{ArmendarizPicon:2007iv}. Since narrow parametric resonance is only present for small values of the coupling where the Universe does not completely preheat, we do not consider it for the remainder of this work.
\begin{figure}
\centering
\includegraphics[height=0.35\columnwidth]{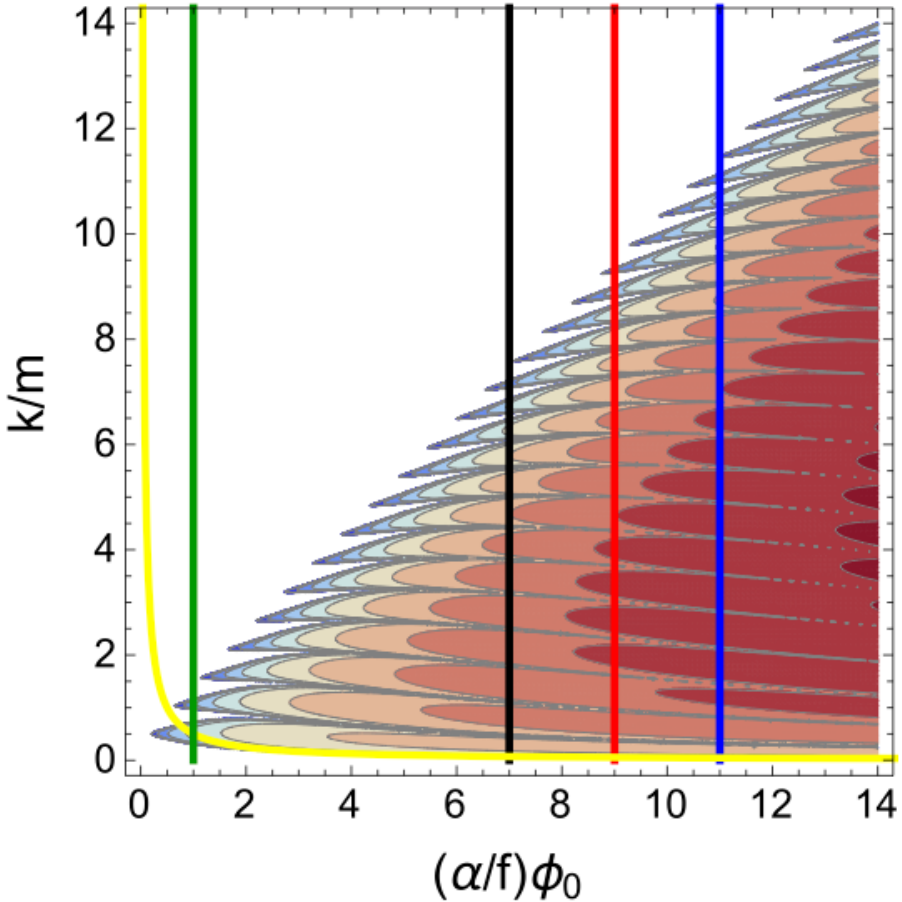} \includegraphics[height=0.35\columnwidth]{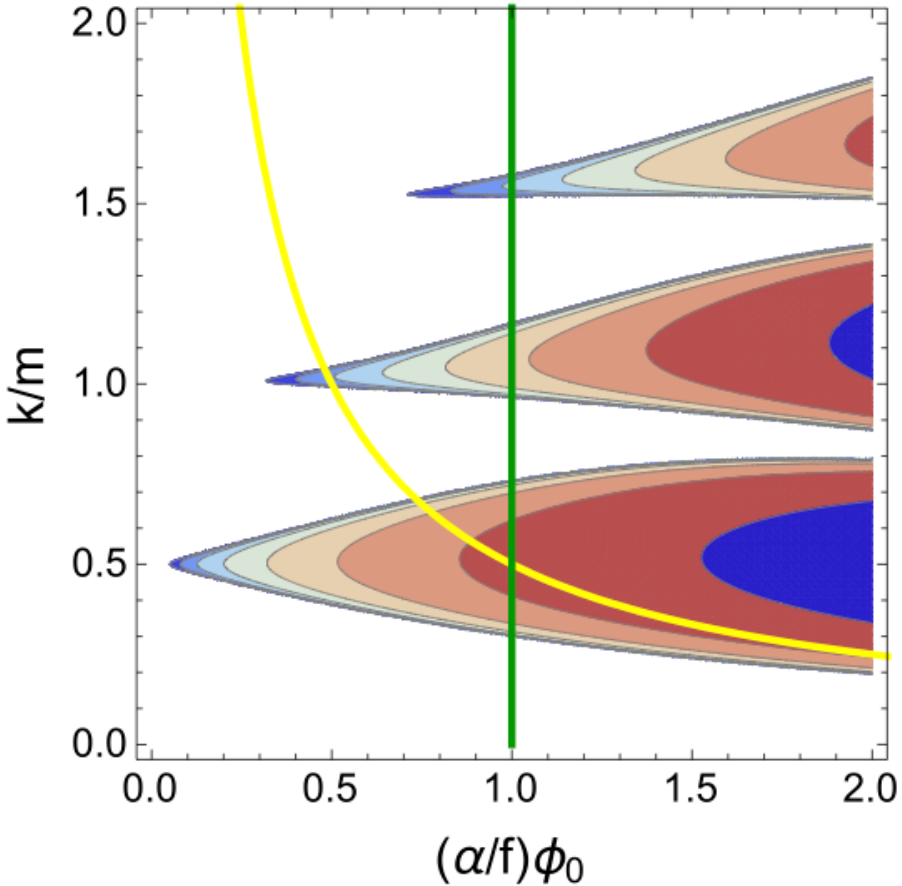} 
\caption{The left panel shows the instability diagram for our process, reparameterized in terms of $k/m$ and $ \phi_0 \alpha / f$. The blue, red and black lines correspond to the same couplings as in Fig.~\ref{fig:matinstab}. The yellow curve divides the regions of narrow and broad resonance defined by $q=1$ and the green line corresponds to $(\alpha/f)\phi_0=1$, which is the value used in \cite{ArmendarizPicon:2007iv}. The right panel shows the instability chart in the region of small coupling $(\alpha/f)\phi_0$ and small $k/m$. Note that the color scale is not the same in both plots. In the left plot, the largest value of the imaginary part of the Mathieu exponent is $\Im(\mu) = 4.2$ while in the right plot it is  $\Im(\mu) = 0.45$. }\label{fig:mathieu_new}
\end{figure}

Previous studies of reheating after axion inflation inflation through gauge-field production, such as \cite{ArmendarizPicon:2007iv}, focused on the region where $q \lesssim 1$, and seem to have missed the efficient, tachyonic preheating phenomenon on which we focus here. As can be seen in Fig.~\ref{fig:mathieu_new}, both the range of wavenumbers, as well as the size of the Floquet exponent are much larger for the range of couplings we study.

These tachyonic instabilities exist for long wavelength modes of the gauge field, but the assumption of a homogeneous inflaton, $\phi (t)$, breaks down quickly once back-reaction occurs.  To see whether rapid production of gauge field quanta occurs, and/or whether this final state is polarized, quickly becomes a numerical question. However, there is some progress to be made using semi-analytic methods by reintroducing the expansion of the Universe into the equation of motion for the gauge fields, as explained in the next section. 

%%%%%%%%%%%%%%%%%%%%%%%%%%%%%%%%%%%%%%%
%%%%%%%%%%%%%%%%%%%%%%%%%%%%%%%%%%%%%%%

\subsection{Semi-analytic treatment}\label{sec:analytics}

\begin{figure}[h!] 
\centering
\includegraphics[width=0.45\columnwidth]{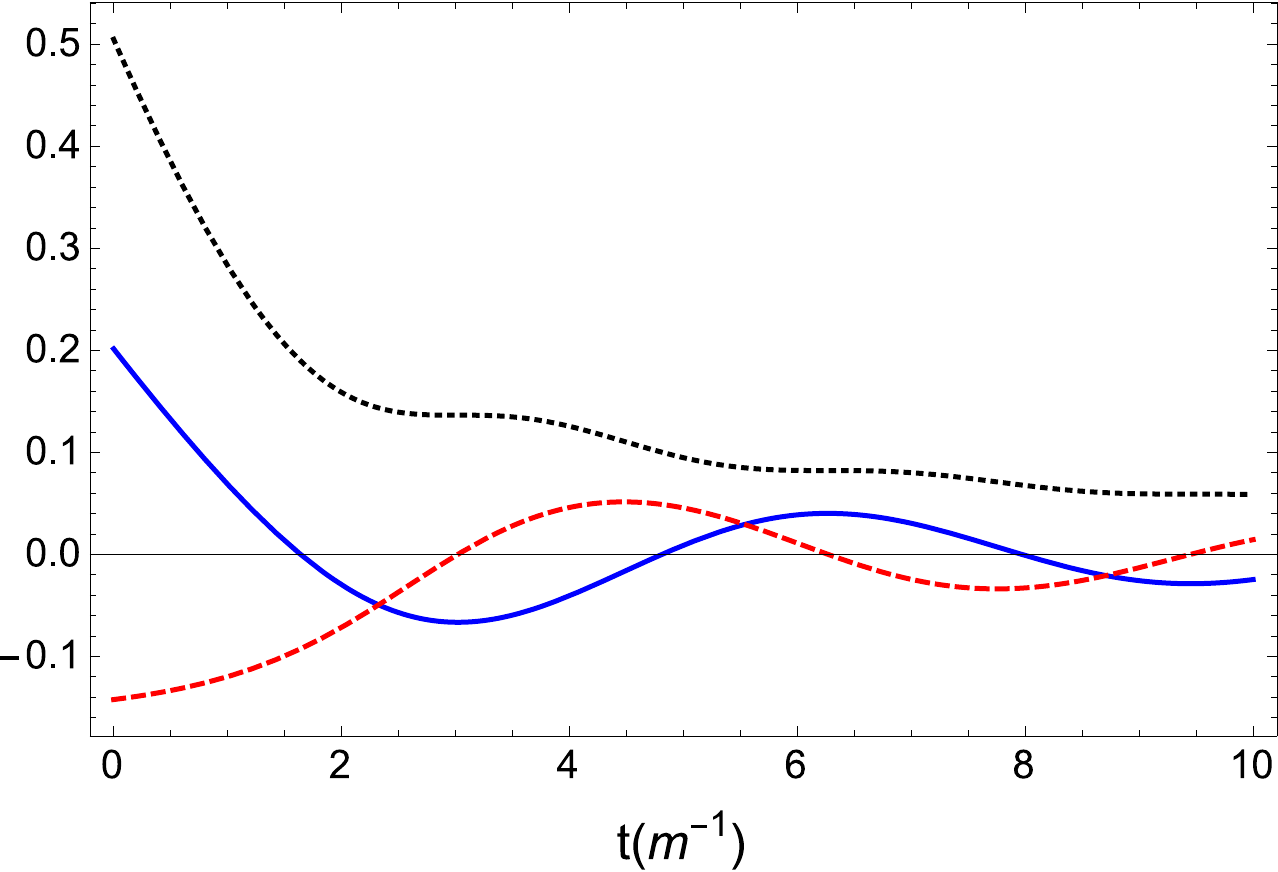}
\caption{The evolution of the axion field $\phi/ m_{\rm pl}$ (blue solid), the velocity $\dot \phi/ m_{\rm pl}$ (red dashed) and the Hubble parameter $H=\dot{a}/a$ (black dotted) after the end of inflation, $t = 0$. Time and related parameters are measured in units of the axion mass $m$.  }
\label{fig:phi_evol}
\end{figure}

A detailed description of tachyonic resonance in the static-universe approximation can be found in \cite{Dufaux:2006ee}, where it is shown that the agreement between these analytic results and numerical simulations (neglecting re-scattering and back reaction) is excellent (for $q\gg1$ and $A_k < 2q-2\sqrt q$). In our current study, we go beyond this approximation.  The richest phenomenology comes from the period between the end of inflation and the first few oscillations of the inflaton. During this epoch, the scale-factor is evolving nontrivially in time and cannot be approximated by an exponential, as it could during slow-roll inflation, or a power-law, as it is in a radiation-dominated universe after reheating. In order to proceed with a semi-analytical treatment, as an approximation,  we model the evolution of the Universe during this time by the evolution of the classical axion in a background FRW spacetime, neglecting the effect of back reaction of the gauge fields on the expansion.  In Fig.~\ref{fig:phi_evol} we plot the evolution of the inflaton field, its velocity, and the Hubble parameter in units of the inflaton mass $m$ for simple chaotic inflation. Without loss of generality, consistent with Section~\ref{sec:duringinflation}, we chose the  value of the axion to be positive during inflation, $\phi>0$, and  its time derivative to be negative, $\dot \phi <0$ (note that here $V_{,\phi} > 0$).  This then determines the signs of these two quantities immediately at the end of inflation.  Note that during the first oscillation the amplitude falls by about a factor of two, and thus approximating the inflaton background by a sinusoidal function with a constant amplitude is insufficient. 

We need to extend the method of \cite{Dufaux:2006ee} in order to treat the case of a time-dependent but non-harmonic effective frequency. We are most interested in the cases where the growth of the gauge field modes is due to tachyonic effects near the end of inflation. In these cases, preheating is extremely efficient, and we focus our attention on the growth of the gauge field modes during the first oscillation of the axion. Due to the strong growth of the fluctuations, analytical methods quickly fail and we enter the regime of strong back-reaction, dominated by non-linear physics. In this regime we  have to rely on full numerical simulations, which we present in the following section. Extending the (linear) analysis beyond the first axion oscillation is straightforward, and can be achieved by combining the procedure we  describe here with the method of \cite{Dufaux:2006ee}.

We start from Eq.~\ref{eqn:kspacegfeqn} written in cosmic time
\beq
\ddot A_k ^\pm + H \dot A_k ^\pm + \left [ \left ({ k\over a} \right )^2 \mp {\alpha \over f} {k\over a} \dot \phi \right ] A^\pm=0,
\eeq
and redefine the gauge field, $\chi^\pm = a^{1/ 2} A^\pm$
\begin{equation}
\ddot \chi_k ^\pm  + \left [ \left ({ k^2\over a^2}  + {\dot a^2 \over 4a^2}  - {\ddot a \over 2a} \right ) \mp {\alpha \over f} {k\over a} \dot \phi \right ] \chi_k^\pm=0,
\end{equation}
which leads to the equation of motion for this rescaled gauge field
\begin{equation}
 \ddot \chi_k ^\pm   + \left [ \omega_k ^\pm (t)  \right ] ^2 \chi_k ^\pm=0.
\label{eq:eomwkb}
\end{equation}

We use the Wentzel-Kramers-Brillouin (WKB) approximation to describe the solution to Eq.~\ref{eq:eomwkb} in the regions where the frequency 
\begin{equation}
 \omega_k ^\pm (t)   = \sqrt{\left ({ k^2\over a^2}  + {\dot a^2 \over 4a^2}  - {\ddot a \over 2a} \right ) \mp {\alpha \over f} {k\over a} \dot \phi},
 \end{equation}
 is varying slowly, that is,  $\left | \dot \omega / \omega^2 \right | \ll 1$.
Following the general procedure of the WKB approximation, we distinguish three regimes based on the behavior of the effective frequency.  We track a mode as it goes from having a positive frequency-squared, $[\omega_k ^\pm (t)]^2$, (regime I) to a negative frequency-squared (regime II), and then back to a positive frequency-squared (regime III). At the interface of these regions, the frequency vanishes ($\omega(t)=0$), and the condition $\left | \dot \omega / \omega^2 \right | \ll 1$ is maximally violated. In practice, this condition is violated for some time interval around the points where the frequency-squared changes sign. 

\begin{figure}[h!] 
\centering
\includegraphics[width=0.45\columnwidth]{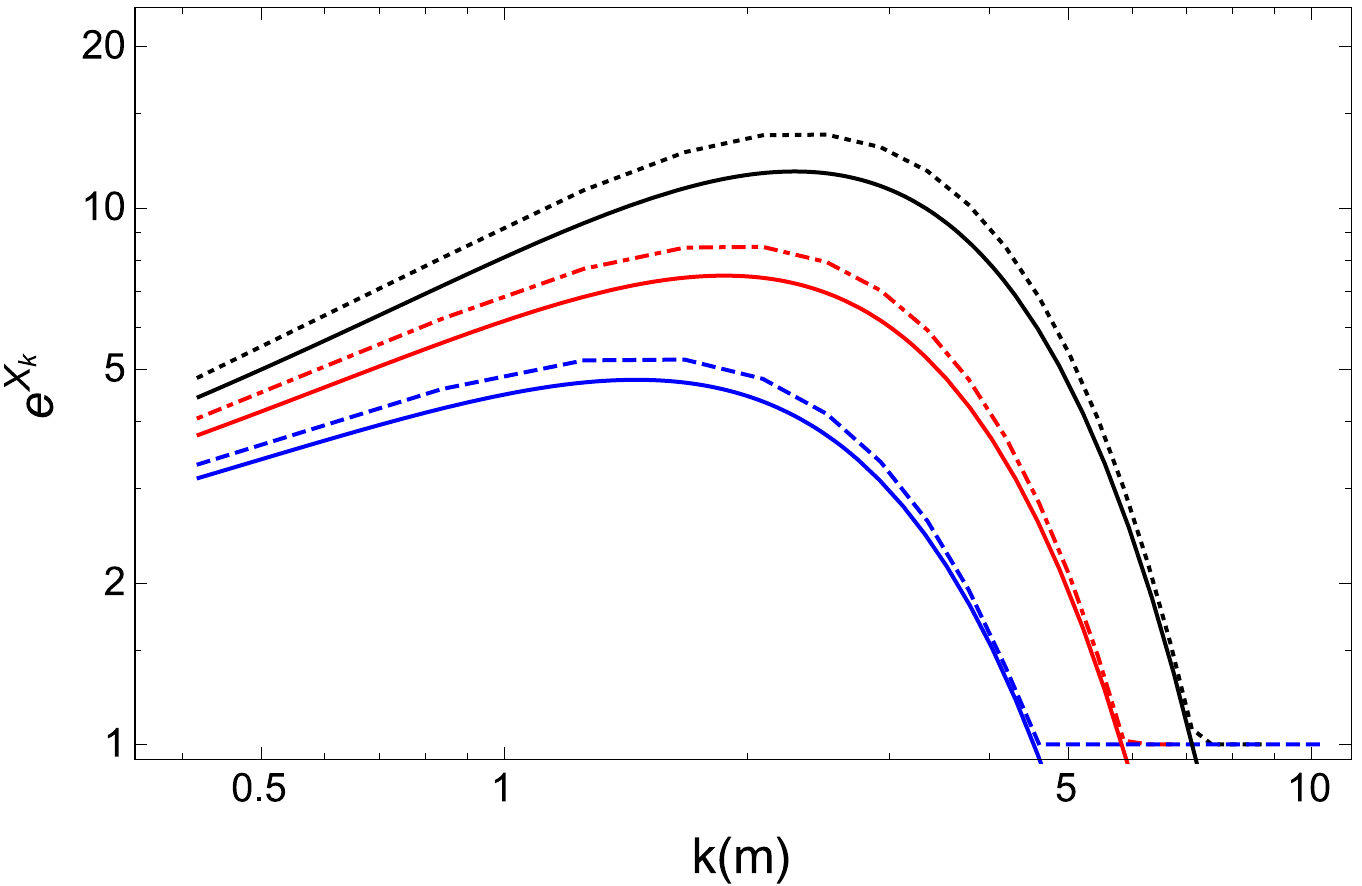}\includegraphics[width=0.45\columnwidth]{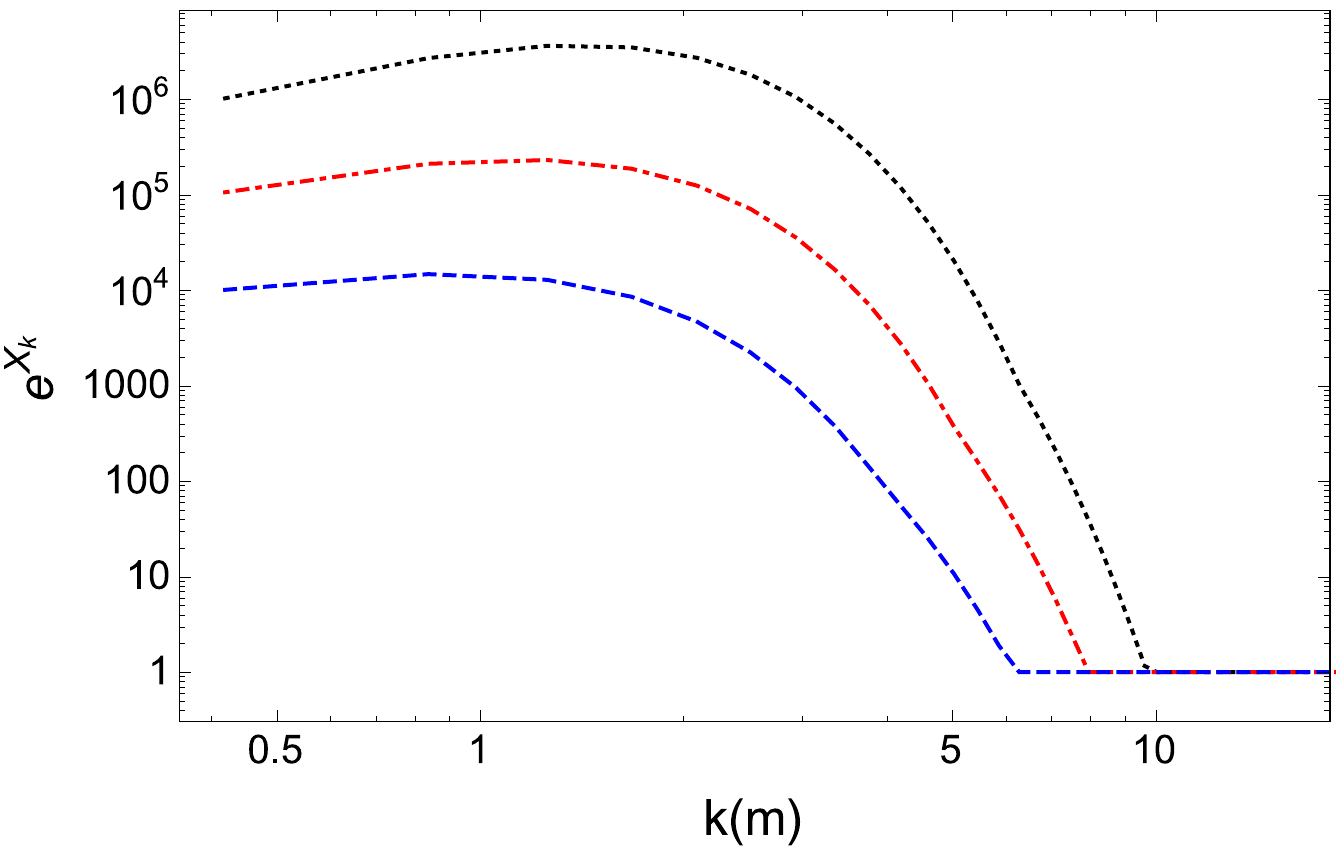}
\caption{The growth factor $X_k$ versus the comoving wavenumber after the first tachyonic instability phase for $A^+$ (left) and $A^-$ (right) based on the semi-analytic calculation. The dotted lines correspond to our WKB formulas, while the continuous lines on the left plot correspond to a modified static-universe approximation. The different lines correspond to different couplings $\alpha / f=35\, m_{\rm pl}^{-1}$ (blue dashed), $\alpha / f=45\, m_{\rm pl}^{-1}$ (red dot-dashed) and $\alpha / f=55\, m_{\rm pl}^{-1}$ (black dotted) for both plots.}
\label{fig:WKB_results}
\end{figure}

Assuming that the frequency is slowly varying, we begin by writing the lowest-order WKB approximation to the solution of Eq.~\ref{eq:eomwkb} in each region \cite{1978amms.book.....B}
\begin{eqnarray}
\chi_k^{\rm I}(t) &=& {\alpha_0\over \sqrt{2\omega_k(t) }} \exp\left ( - i \int_{t_0}^t \omega_k(t')dt' \right )  + {\beta_0 \over \sqrt{2\omega_k(t) }} \exp\left (  i \int_{t_0}^t \omega_k(t')dt' \right ) ,
\\
\chi_k^{\rm II}(t) &=&{a\over \sqrt{2\Omega_k(t) }} \exp\left ( - \int_{t_1}^{t} \Omega_k(t')dt' \right )  +{b\over \sqrt{2\Omega_k(t) }} \exp\left (  \int_{t_1}^{t} \Omega_k(t')dt' \right ) ,
\\
\chi_k^{\rm III}(t) &=& {\alpha\over \sqrt{2\omega_k(t) }} \exp\left ( - i \int_{t_0}^t \omega_k(t')dt' \right ) + {\beta \over \sqrt{2\omega_k(t) }} \exp\left (  i \int_{t_0}^t \omega_k(t')dt' \right ) ,
\label{WKBeq3}
\end{eqnarray}
where $\Omega^2(t) = -\omega^2(t)$, and $\omega_k(t_1) = \omega_k(t_2)=0$ and the integrals in the expression for $\chi_k^{III}(t)$ are performed for the region where $\omega^2(t)>0$.  
We are ultimately interested in the two coefficients $\alpha$ and $\beta$ which describe the amplitude of the mode after its brief growth due to the tachyonic instability.
In regime I we match the solution to the asymptotic past (Bunch-Davies vacuum), giving us $\alpha_0=1$ and $\beta_0=0$. After matching at the two interfaces between regimes I and II and regimes II and III (by Taylor-expanding $\omega^2(t)$ near the turning points and using the asymptotic form of the Airy functions) the coefficients are
\begin{equation}
\alpha = e^{X_k},
\quad {\rm and } \quad
\beta = -ie^{X_k}e^{-2i\theta_k}, 
\end{equation}
where 
\begin{equation}
X_k = \int_{t_1}^{t_2} \Omega_k(t')dt'
,\quad {\rm and} \quad
\theta_k = \int_{t_0}^{t_1} \omega_k(t')dt'.
\end{equation}
Before proceeding, we note that the WKB approximation is accurate only in the broad resonance regime, $q\gg1$. This translates to the condition that 
\begin{equation}
\frac{k}{am}\frac{\alpha}{f} \phi_0 \gg 1.
\end{equation} For typical values of the coupling and the inflaton amplitude, this condition restricts the validity of the WKB approximation to $k \gtrsim m$. As shown in Fig.~\ref{fig:WKB_results}, the maximum amplification occurs well within the region of validity of our WKB calculation, especially for higher couplings.

We study each polarization individually, starting with the mode $A^+$. In the absence of back reaction, the axion velocity, $\dot\phi$, takes its maximum positive value at time $(t_* -t_0 )\approx 4.5 \,m^{-1}$ after inflation ends,\footnote{In this section we denote by a subscript $0$ the value of a quantity at the end of the inflationary phase where $\epsilon_H \equiv -\dot{H}/H^2 = 1$ for the first time.} which sets the interval during which $A^+$ is tachyonic for the first time. The evolution of the inflaton field $\phi(t)$ is close to sinusoidal at this stage, meaning that we can use a modified static-universe approach. However, we employ our full expanding-universe WKB-method as a way of testing its accuracy and providing a unified treatment of the two gauge-field polarizations. 

The evolution of the second polarization, $A^-$, is more involved due to the fact that regime II, characterized by $\omega^2<0$, starts while the Universe is still inflating and continues through the end of inflation into the first oscillation of the inflaton. This complication does not significantly alter our analysis. To proceed, we simply need to initialize the mode in regime I, sufficiently early during the inflationary stage, before its tachyonic transition and follow it, using the WKB approximation, through this transition and the end of inflation. The final formulas are exactly the same in this case.

\begin{figure}[h!] 
\centering
\includegraphics[width=0.45\columnwidth]{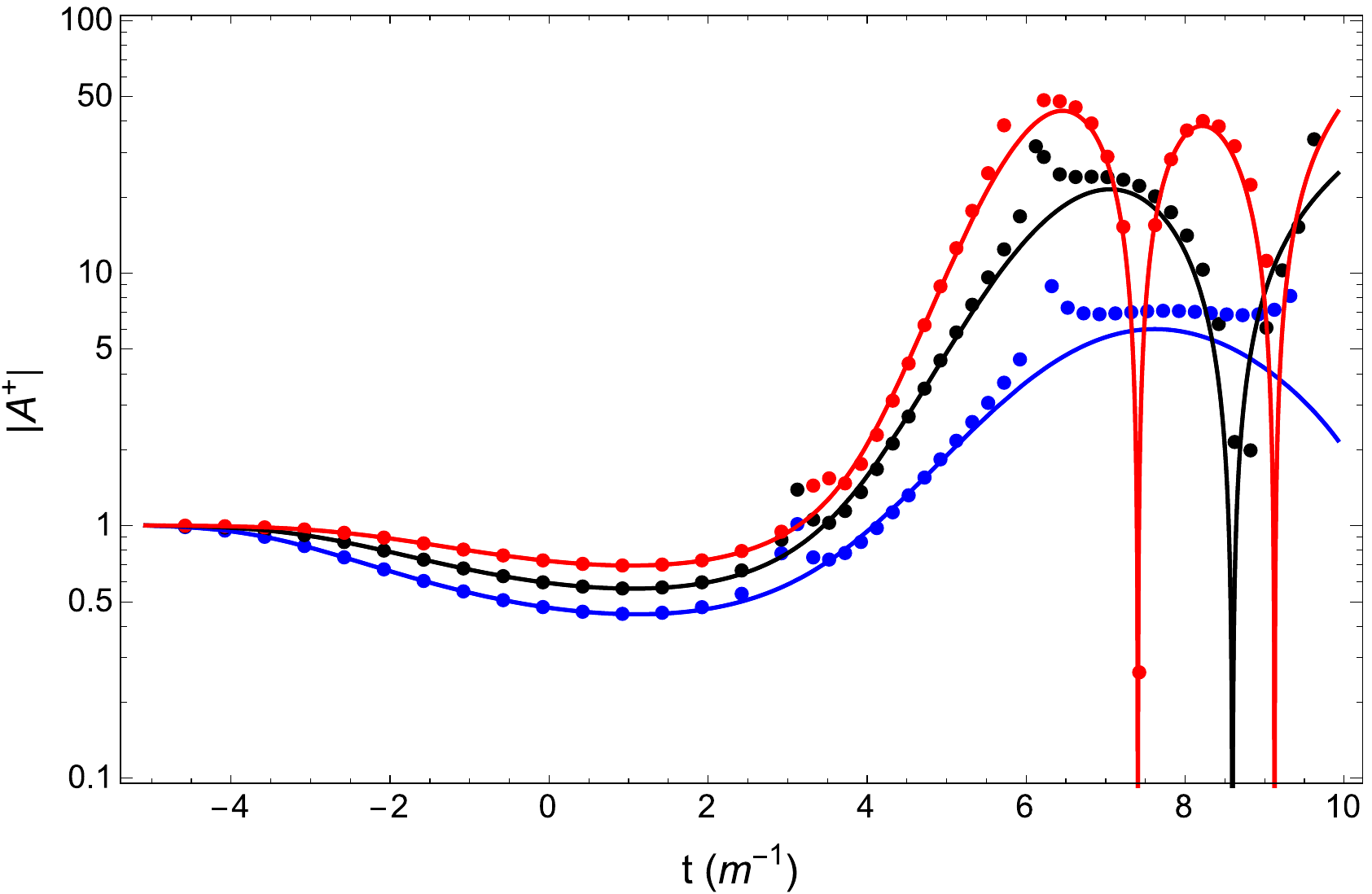}\includegraphics[width=0.45\columnwidth]{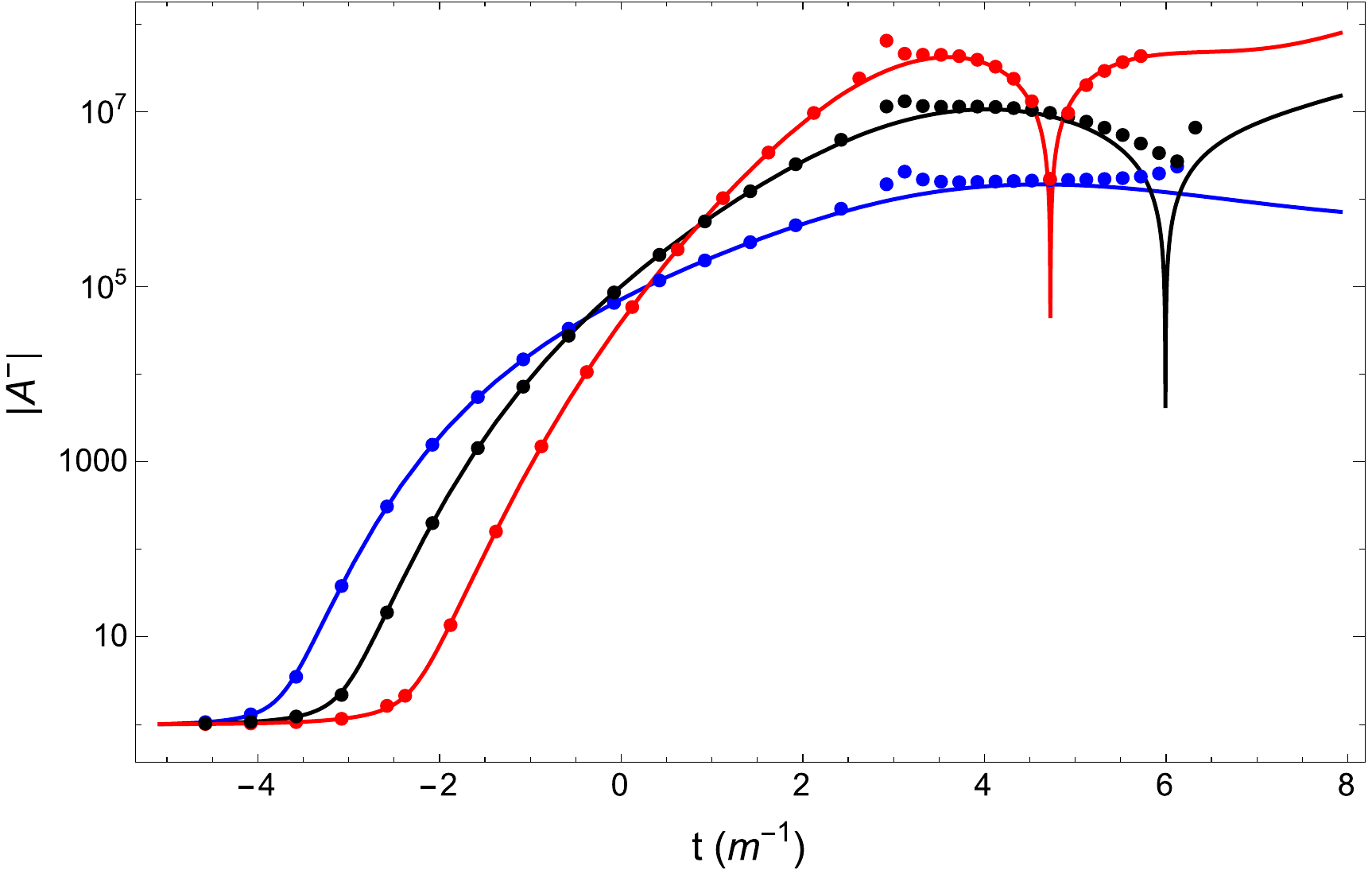}
\caption{The amplification of gauge fields during the first tachyonic instability phase for $A^+$ (left) and $A^-$ (right) based on numerical simulations and the semi-analytic calculation (for $\alpha/f=70\,m_{\rm pl}^{-1}$). On the left panel the lines and dots correspond to numerical results for wavenumbers $k/(aH)=1 $(blue), $k/(aH)=e$ (black) and $k=k/(aH)=e^2$ (red). On the right panel the lines and dots correspond to numerical results for wavenumbers $k/(aH)=e^{-1} $ (blue), $k/(aH)=1 $ (black) and $k/(aH)=e$ (red).  In all cases $aH$ is evaluated at the end of inflation, and $t = 0$ corresponds to the end of inflation.} 
\label{fig:WKB_test}
\end{figure}

We can evaluate the validity and accuracy of the semi-analytical method described above by comparing the results directly with a full numerical solution of the linear equations of motion. Using {\sc Mathematica}, we solve the homogeneous Klein-Gordon and Friedman equations for the evolution of the background axion and spacetime. On this background, we follow three gauge-field modes for $A^+_k$ whose wavelengths are equal to the horizon, and one and two e-foldings smaller than the horizon at the end of inflation, i.e.\ $k/(aH) =\left\{1,\,e, \, e^2\right\}$ respectively at the end of inflation.  We also track three modes for $A^-_k$, one that exits the horizon one e-fold before the end of inflation, a mode whose wavelength is equal to  the horizon, and a third that is one e-fold smaller than the horizon at the end of inflation, i.e.\ $k/(aH) =\left\{ e^{-1}, \,1,\,e\right\}$ respectively at the end of inflation. In order to facilitate the comparison, we take the mode amplitudes to have unit size at the start of our simulation. The WKB condition $\left | \dot{ \omega} / \omega^2\right | \ll 1$ is violated around the points where $\omega=0$ for each mode and $\left | \dot{ \omega} / \omega^2\right | \le {\cal O} (0.1)$  during the tachyonic regime. This limits the accuracy of the approximation. The approximation could be improved by making use of a transformation of variables similar to \cite{Martin:2002vn}, however, given the good agreement with the numerical results shown in Fig.~\ref{fig:WKB_test}, and the fact that we are trying to understand the results of full lattice simulations rather than substitute them, we do not attempt to refine the WKB method used here.

In Fig.~\ref{fig:WKB_test} we show the excellent overall agreement between the numerical results from direct numerical solution in {\sc Mathematica} and those obtained from the WKB approximation. As expected the approximation diverges at the points where $\omega^2(t)=0$, and closely follows the curves as one moves away from these points. Furthermore, we can see that for both $A^+$ and $A^-$ the accuracy of the approximations decreases with decreasing wavenumber $k$ (leading to decreasing $q$), especially after the tachyonic transition. Although we are pushing the limits of the WKB approximation, the accuracy of the resulting growth rates is remarkably robust.

These results demonstrate that this semi-analytical method can be used to accurately estimate the growth of the gauge fields during the first inflaton oscillation, if one neglects rescattering and back-reaction effects. In Fig.~\ref{fig:WKB_results}, we plot the growth factor $X_k$, which shows by how much the amplitude of each gauge mode has grown after the first tachyonic regime. Note that this WKB method breaks down at small wavenumbers. 

For the mode $A^+$ we can make a comparison to the static-universe calculation by using rescaled parameters. 
The amplitude of the axion oscillations has decreased due to the expansion of the Universe after the end of inflation (as shown in Fig.~\ref{fig:phi_evol}). For the tachyonic half-period of interest, the behavior of the axion field is well approximated by 
\begin{equation}
\dot \phi (t) \approx -0.05 m\, m_{\rm Pl} \cos(mt+\Delta \theta),
\end{equation}
where $\Delta \theta$ is a phase offset that allows the time of zero-crossing to correspond to our model. The wavenumbers of the modes under consideration are also redshifted. The modes used for our full WKB calculation were measured in terms of the scale factor at the end of inflation $a_0$, while the Universe has grown by a factor of $a/a_0\approx 2.6$ by the middle of the first tachyonic regime for $A^+$. We thus rescale the wavenumber used in our static-universe calculation by $2.6$.  As we can see in Fig.~\ref{fig:WKB_results} the results are very close, giving us a simple physical way to understand the result of using the WKB method in an expanding universe. 

The polarization $A^-$ is more complicated due to the fact that it becomes tachyonic during the inflationary phase. From Fig.~\ref{fig:WKB_results} note that the growth factors for a given mode are significantly larger, and a larger range of wavenumbers get amplified.  
%Furthermore,  a small ``knee'' is present in all curves at comoving wavelengths $k/m\approx6.8,~11.4,~15.8$, and $20.6$ for $\alpha/f =30\, m_{\rm pl}^{-1},~50\, m_{\rm pl}^{-1},~70\, m_{\rm pl}^{-1}$ and $90\, m_{\rm pl}^{-1}$ respectively. The modes with wavenumbers above this knee become tachyonic after the end of inflation, resulting in a different shape of $-\omega^2(t)$ when evaluated during the tachyonic phase. 
We do not compare this case with a static-universe approximation, since $\dot \phi$ is not well approximated by a sinusoidal function in the regime where $A^-$ is tachyonic, even after inflation has ended. This is due to the Hubble friction term, as shown in Fig.~\ref{fig:phi_evol}.

There is one further complication we need to keep in mind when comparing the semi-analytic predictions of this section with the full lattice simulations presented in Section \ref{nonlinear}. Once the coupling to the gauge fields is turned on, these fields begin to backreact. This causes inflation to end at a slightly different value of $\phi$ for each coupling $\alpha /f$, as shown in Table \ref{tab:phi0}. We also show the values of the field and field velocity two e-foldings before inflation ends. Note that these values are all close to the zero coupling case, which indicates that backreaction has not yet become important yet.The WKB analysis of this section was performed using the evolution of the axion in the limit when we neglect back reaction from the gauge fields, equivalently in the limit of $\alpha /f \to 0$.  
\begin{table}[h]
\begin{center}
\begin{tabular}{|c|c|c|c|c|}\hline
$m_{\rm pl}\alpha/f $ & $\phi_{\rm e^{-2}}$ & $\dot{\phi}_{\rm e^{-2}}/m $& $\phi_{\rm end}$ & $\dot{\phi}_{\rm end}/m$ \\\hline
$0 $ & $0.563$ & $-0.158$ &$0.201 $&  $-0.143$  \\
$30 $ & $0.563$ & $-0.158$ &$ 0.201 $&$ -0.143 $\\
$35 $ &$0.563$ & $-0.158$& $0.201 $&$ -0.143 $ \\
$40 $ & $0.564$ &$-0.158$ & $0.203 $& $-0.142 $ \\
$45 $ & $0.571$ &$-0.158$ & $0.217$ & $-0.142 $ \\
$50$  & $0.580$  &$-0.158$ & $0.237$ & $-0.147 $ \\
$55$  &$0.579$ & $-0.158$ &$ 0.239 $&$ -0.132 $\\
 $60$ &$0.572$ & $-0.158$&$ 0.235$ &$ -0.115$\\
 $65$  & $0.564$& $-0.158$ & $0.231$ &$-0.102 $
\\ \hline
\end{tabular}
\end{center}
\caption{Field conditions two e-folds before the end of inflation and at the end of inflation  for $V = m^2\phi^2/2$ and $m/m_{\rm pl} = 1.06 \times 10^{-6}$ including the gauge-field backreaction. Note that $\phi_{\rm end}$ does not monotonically increase with the coupling. This is likely an  artifact of the approximation. \label{tab:phi0}}
\end{table}%

%\begin{table}[h]
%\begin{center}
%\begin{tabular}{|c|c|c|c|c|c|}\hline
%$m_{\rm pl}\alpha/f $ & $\phi_{\rm end}$ & $\dot{\phi}_{\rm end}/m $& $m_{\rm pl}\alpha/f$ & $\phi_{\rm end}$ & $\dot{\phi}_{\rm end}/m$ \\\hline
%$0 $&$0.201 $&  $-0.152$ & $55$ &$ 0.239 $&$ -0.132 $\\
%$30 $&$ 0.201 $&$ -0.152 $ & $60$ &$ 0.235$ &$ -0.115$\\
%$35 $& $0.201 $&$ -0.152 $& $65$ & $0.231$ &$-0.102 $\\
%$40 $& $0.203 $& $-0.152 $& $70$ &$0.229 $& $-0.0913 $\\
%$45 $& $0.217$ & $-0.151 $& $75$ & $0.229$ & $-0.0823$\\
%$50$ & $0.237$ & $-0.147 $& $80$ & $0.230$& $-0.0750$\\ \hline
%\end{tabular}
%\end{center}
%\caption{Field conditions at the end of inflation including back reaction for $V = m^2\phi^2/2$ and $m/m_{\rm pl} = 1.06 \times 10^{-6}$. Note that $\phi_{\rm end}$ does not monotonically increase with the coupling. This is likely an  artifact of the approximation. \label{tab:phi0}}
%\end{table}%

%%%%%%%%%%%%%%%%%%%%%%%%%%%%%%%%%%%%%%%
%%%%%%%%%%%%%%%%%%%%%%%%%%%%%%%%%%%%%%%
%%%%%%%%%%%%%%%%%%%%%%%%%%%%%%%%%%%%%%%

\section{Non-linear lattice simulations}
\label{nonlinear}

The results of Section~\ref{sec:analytics} imply that at the couplings of interest, the dynamics of the axion-gauge field system very quickly become highly non-linear and require numerical analysis.  Numerical methods that evolve scalar fields in an expanding background have been around for a couple of decades. Many numerical methods have been developed \cite{Felder:2000hq,Frolov:2008hy,Easther:2010qz,Huang:2011gf}, and all have regimes in which they are most successful.  Here we use {\sc GABE} \cite{GABE} (as in \cite{Deskins:2013dwa}), where the axion and the gauge field are defined on a discrete lattice (grid) with $256^3$ points.  The software uses a second-order Runge-Kutta integration method to solve Eqs.~\ref{axioneom}, \ref{gaugeeom1} and \ref{gaugeeom2} alongside the self consistent expansion of space-time Eq.~\ref{ffriedman}. We work in Lorenz gauge, $\partial^\mu A_\mu = 0$, to evolve the gauge fields. In this gauge, the Gauss law constraint becomes a dynamical equation for $A_0$ which we solve in parallel with Eqn.\ \ref{gaugeeom2}.  Note that, although we initialize our fields using solutions of the linear equations of motion in Coulomb gauge, these gauge fields are equivalent to the fields in Lorenz gauge. To see this, note that if one begins with fields in Coulomb gauge, and performs a gauge transformation to Lorenz gauge $A_{\mu}^{\rm Lorenz} = A_{\mu}^{\rm Coulomb} - \partial_{\mu}\alpha$, the Lorenz gauge condition ($\partial^{\mu}A^{\rm Lorenz}_{\mu} = 0$) becomes $(\partial_\tau^2 - \partial_i\partial_i)\alpha = 0$, which is simply the residual gauge symmetry of the Lorenz gauge.  Unless otherwise noted, all simulations use a box size that is $L=15\,{m^{-1}}$ at the end of inflation and are run using the parallel processing standard {\sc OpenMP} with 12 threads.  The equations of motion for the gauge degrees of freedom, $A_\mu$, as well as the axion, $\phi$, are integrated using a second-order (midpoint method) Runge-Kutta integration scheme.  We normalize quantities so that wave numbers are expressed in terms of the size of the box at the end of inflation, $L=15\,m^{-1}$, and $a=1$ (at the end of inflation).

\begin{figure}[h!] 
\centering
\includegraphics[width=0.45\columnwidth]{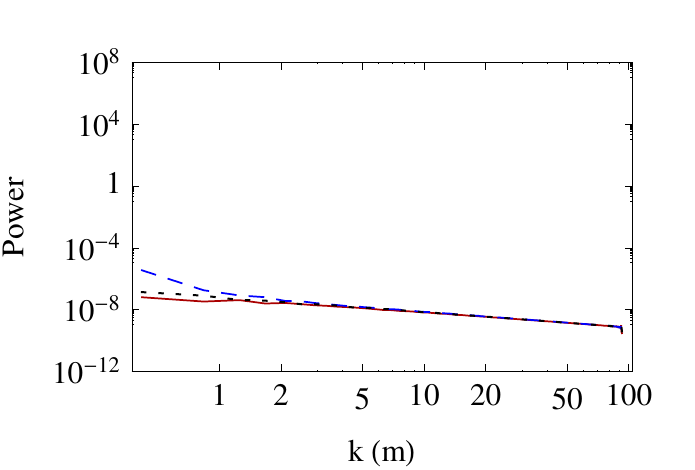}\includegraphics[width=0.45\columnwidth]{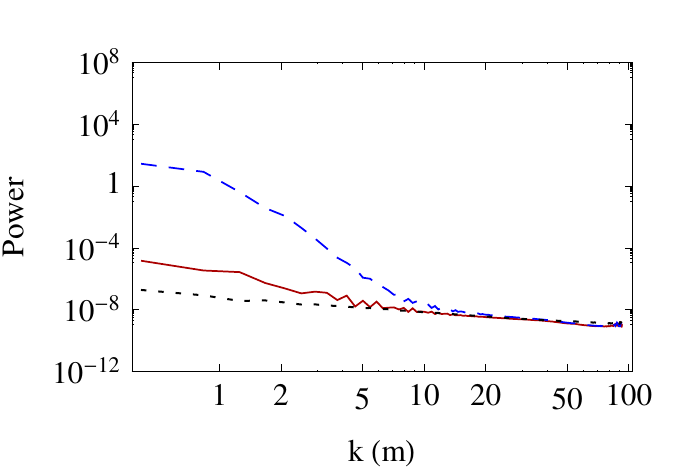}\\
\includegraphics[width=0.45\columnwidth]{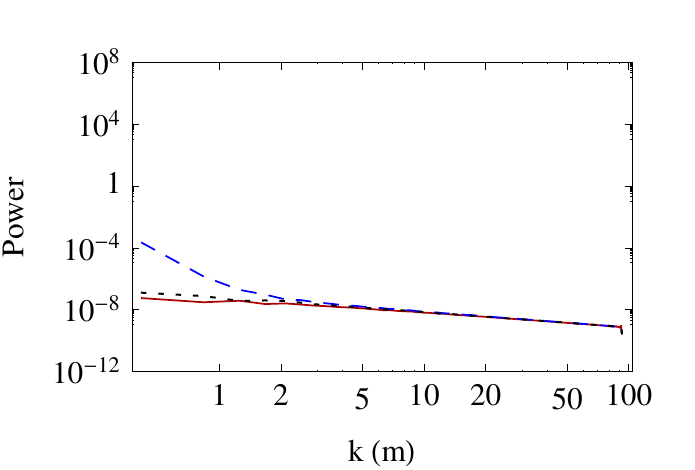}\includegraphics[width=0.45\columnwidth]{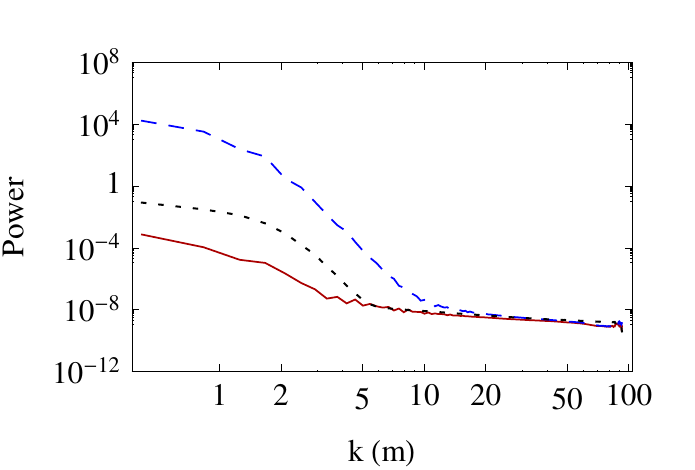}
\caption{Upper panels: The initial power spectrum of the fields, $A_+$ (red, solid) $A_-$ (blue, dashed) and $\phi$ (black, dotted) at $\alpha / f =45\,m_{\rm pl}^{-1}$.  
The left panel shows the spectra at the beginning of the simulation which occurs at two-efoldings before the end of inflation, $a=e^{-2}$. The right panel shows the spectra of the same three fields at the end of inflation.  The lower panels show the same spectra at a coupling of $\alpha / f =65\,m_{\rm pl}^{-1}$, the highest coupling that we simulate.  Note the long-wavelength fluctuations of the axion generated during the end of inflation by for the largest couplings.
}
\label{fig:initmodes}
\end{figure}

We begin our simulations two e-foldings before inflation ends, defining this point as $\tau = 0$. As described at the end of Section \ref{sec:duringinflation}, we determine the value of the homogeneous field and its derivative by numerically evolving (using {\sc Mathematica}) the system of Eqs.~\ref{eqn:backreactfried} - \ref{eqn:backreactKG} together with the approximations of Eq.~\ref{eqn:GFenergy}.  At this point the box-size is $L_0 = L e^{-2} \approx 2 m^{-1}$ which is just larger than the Hubble Scale, $H^{-1}  = \sqrt{3/8\pi}\rho^{-1/2} \approx 1.2 m^{-1}$, where the final approximation varies slightly from coupling to coupling.  We initialize the power in the $A_{\pm}$ modes by numerically evolving Eq.~\ref{eqn:kspacegfeqn} for each physical mode, tracking it from when it was well inside the horizon ($\tau \rightarrow -\infty$) until two e-foldings before the end of inflation.  Since we have no analytic form for the fluctuations of $\phi$ at this time, we initialize it in the Bunch-Davies vacuum, $\left<\phi_k^2\right> = 1/\sqrt{2\omega}$, where it is an excellent approximation due to the fact that almost all of our modes are sub-horizon.  Using this procedure, the modes that leave the horizon during the final two e-foldings of inflation are done so self-consistently and the spectrum of perturbations for $\phi$ is consistent with our equations of motion. Fig.~\ref{fig:initmodes} shows the spectra at the beginning of the simulation and the end of inflation showing the amplification of large wavelength modes during the final stages of inflation.

\begin{figure}[h!] 
\centering
\includegraphics[width=.45\columnwidth]{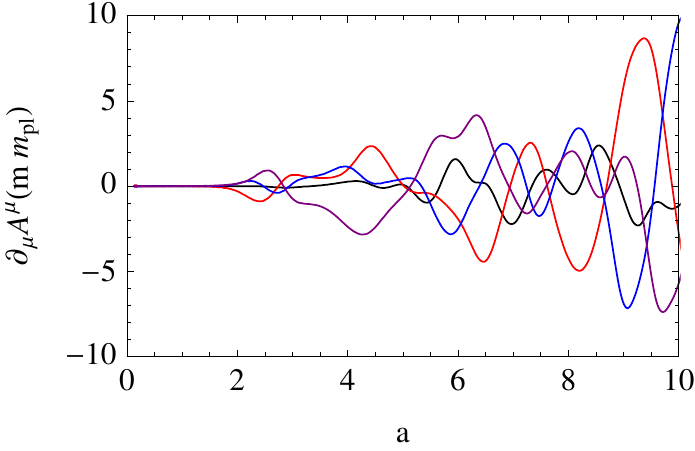}\phantom{sp}\includegraphics[width=.45\columnwidth]{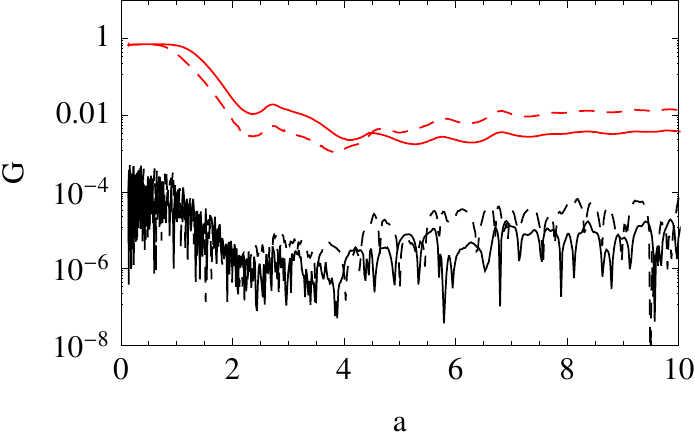}
\caption{The left panel shows the four terms of $\partial_\mu A^\mu$ that contribute to calculating the gauge condition at an arbitrary point in the simulation box: $\partial_0A^0 $ (black), $\partial_1 A^1$ (red), $\partial_2 A^2$ (blue), and $\partial_3 A^3$ (purple).  The right panel shows a plot of $G(\tau)$ averaged over all points on the box, $\left<G\right>$, (black, solid) and the RMS average, $\sqrt{\left<G^2\right>}$, (red, solid), at $N=256$ compared to the same two quantities (same colors but dashed) for $N=128$.  This parameterizes how well the gauge condition is satisfied and is shown for $ \alpha/f = 45\, m_{\rm pl}^{-1}$}
\label{fig:gaugecheck}
\end{figure}

After we set the initial spectrum of $A^\pm_{{\bf k}}$ in momentum space we project these onto the gauge fields
\begin{equation}
\vec{A}_{\bf k}  = \vec{\epsilon}_+({\bf k}) A^+_k  + \vec{\epsilon}_-({\bf k}) A^-_{k},
\end{equation}
and (inverse) Fourier transform them into configuration space using a set of projection operators, $\epsilon_{ij}^\pm$, that satisfy the relations
\begin{equation}
\label{transversedefofproj}
{\bf k}\cdot \vec{\epsilon}_\pm = 0, \quad
%\end{equation}
%and 
%\begin{equation}
 {\bf k} \times \vec{\epsilon}_{\pm} = \mp ik \vec{\epsilon}_{\mp}.
\end{equation}
These relations set only the spatial components of the gauge field, $\vec{A}({\bf x}, \tau = 0)$, on the initial surface.  Since we are numerically tracking the values of the full four-potential, $A^\mu$, we must check to make sure the Lorenz gauge condition, $\partial^\mu A_\mu = 0$, is obeyed in configuration space as required by our equations of motion. The definition of the polarizations, Eq.~\ref{transversedefofproj}, requires that $\dot{A}^0 = 0$ ($A^0 = {\rm constant}$) on the initial slice.  Therefore any choice of $\vec{A}_\pm$ (with the choice $A_0=0$) obeys the gauge condition.  We are, of course, neglecting any effect that the initial conditions of $A^\mu$ have on $\phi$ on the initial surface. However, we begin our simulations during inflation where all modes of interest are sub-horizon. After the fields are initialized, there is an amplification of modes of $\phi$ that reaches equilibrium well before the first zero-crossing of the field. To make sure that we don't depart from the gauge-condition surface, obeying Lorenz gauge, we track the size of
\begin{equation}
G(\tau) = \frac{\partial_0A_0 - \vec{\nabla}\cdot \vec{A}}{\sqrt{(\partial_0A_0)^2 + (\partial_1A_1)^2 + (\partial_2A_2)^2 + (\partial_3A_3)^2}}
\end{equation}
as in \cite{Deskins:2013dwa}.  Fig.~\ref{fig:gaugecheck} shows a plot of $G(\tau)$ averaged over the box and the RMS averaged value, $\sqrt{\left<G^2\right>}$, for a simulation with  $\alpha/f = 45\,m_{\rm pl}^{-1}$. We compare this at two resolutions, $N=128$ and $N=256$.  The reader should understand this as a proxy for {\sl staying on} the gauge surface---when the simulation diverges from satisfying the gauge condition, this measure quickly grows large.  At early times, since we do not `cutoff' the initial spectra, the power in the very highest frequency modes contribute to variations in the finite-derivatives we use to calculate $G(\tau)$ are artificially important, and contribute to the size of the RMS value, since the field values are so small.  Once the modes of the gauge field are populated, at the end of inflation, this measure is increasingly good.   

The goals of our simulations are to see if the tachyonic and/or parametric instabilities identified in Section \ref{sec:instabandres} lead to the efficient generation of gauge modes, whether the energy deposited in those gauge modes is enough to preheat the Universe, and whether that final state has any anisotropy in the two polarization states of the gauge field, as a naive interpretation of Fig.~\ref{fig:WKB_results} would suggest.  To parameterize the success of the first and second of these questions, we calculate the total energy in the gauge field (as defined in Eq.~\ref{eqn:friednmaneq}) 
\begin{equation}
\rho_{EM} = \frac{1}{2}\left<E^2+B^2\right>,
\label{eq:rhoEM}
\end{equation}
although we express this as a fraction of the total energy density of the Universe, $\rho_{EM}/\rho_{tot}$, for various couplings.

\begin{figure}[h!] 
\centering
\includegraphics[width=0.45\columnwidth]{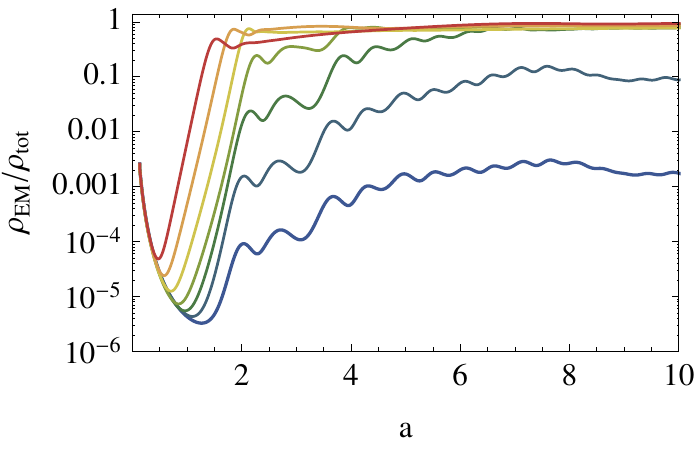}
\includegraphics[width=0.45\columnwidth]{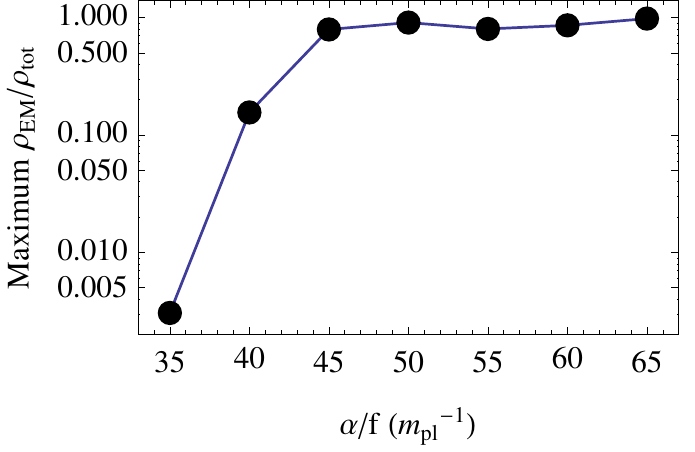}
\caption{The total energy fraction in the gauge field $\rho_{\rm EM}/\rho_{\rm total}$ as a function of time for a variety of couplings.  We probe couplings between $ \alpha/f = 35\,m_{\rm pl}^{-1}$ to $ \alpha/f = 65\,m_{\rm pl}^{-1}$ in increments of $\alpha/f = 5\,m_{\rm pl}^{-1}$.  The left panel shows these couplings as a function of time for each run. The couplings go from largest on the top (red) to the lowest on the bottom (blue) along the rainbow spectrum. The two lowest couplings, $\alpha/f =  35\,m_{\rm pl}^{-1}$ and $40\,m_{\rm pl}^{-1}$ do not completely preheat. The right panel shows the maximum ratio of $\rho_{\rm EM}/\rho_{\rm total}$ for each value of the coupling, $\alpha/f$. The initial energy of the gauge fields is red-shifted away during the last two e-foldings of inflation since we are unable to introduce shorter-wavelength modes during the simulation.}
\label{emendate}
\end{figure}

As the simulations progress, we see that energy is, generically speaking, quickly and efficiently transferred into the gauge fields.  Fig.~\ref{emendate} shows the evolution of the ratio of the energy in the gauge fields to the total energy as a function of time through the simulation for different values of the couplings.
In all cases where $\alpha /f \gtrsim 45\,m_{\rm pl}^{-1}$ the Universe completely preheats and almost all of the energy of the simulation is transferred into the gauge fields.  In most cases, this happens during the first few oscillations of the axion field, justifying the need for full non-linear simulations.  Only in the case of a marginal coupling $\alpha/f = 45\,m_{\rm pl}^{-1}$ does the axion take almost two full oscillations before the gauge field dominates the energy density.  Below, we discuss a few regimes identified by these simulations.

%%%%%%%%%%%%%%%%%%%%%%%%%%%%%%%%%%%%%%%
%%%%%%%%%%%%%%%%%%%%%%%%%%%%%%%%%%%%%%%

\subsection{Early tachyonic resonance}
\label{tachyinstab}

The most significant difference between this model and previous studies of gauge fields during reheating following inflation \cite{Deskins:2013dwa} is the prediction that one of the polarizations of the gauge field is tachyonic each time the homogeneous mode crosses zero.  This should persist as long as the inflaton is (dominantly) coherent and we can treat it as a strictly time-dependent quantity in Eq.~\ref{akeom}.  We are using a box whose physical size increases over the course of the simulation.  The longest wavelength we are able to probe at the beginning of the simulation corresponds to a minimum wavenumber, $k_{\rm min} = 2\pi/L \approx 0.4\, m^{-1}$ (although this does grow during the simulation as the scale factor increases).  

\begin{figure}[h!] 
\centering
\includegraphics[width=0.45\columnwidth]{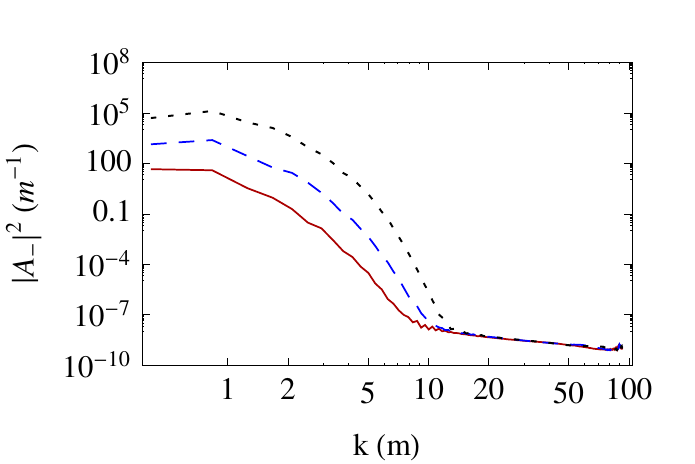}
\includegraphics[width=0.45\columnwidth]{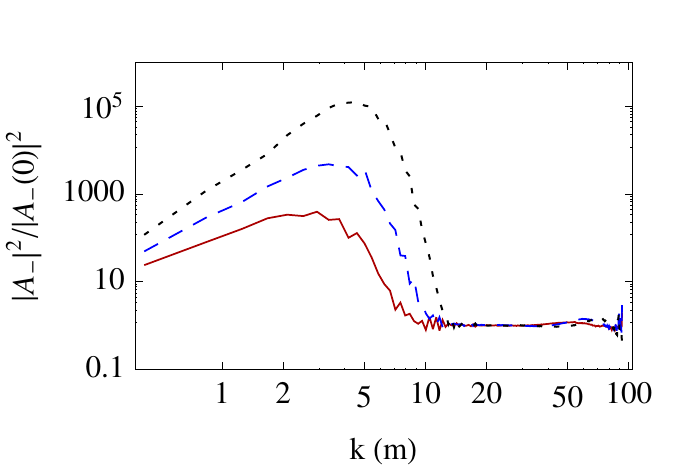}
\caption{We compare the power spectrum of the $A^-$ mode for simulations in which the coupling is $\alpha/f = 35 \,m_{\rm pl}^{-1}$ (red, solid) $\alpha/f = 45\,m_{\rm pl}^{-1}$ (blue, dashed), and $\alpha/f = 55\,m_{\rm pl}^{-1}$ (black, dotted).  The left panel shows the power spectrum of the $A^-$ at the time when the axion crosses zero for the first time for these three couplings.  The right panel shows the ratio of $A^-$ evaluated at the first zero crossing to the initial spectrum, capturing the enhancement of $A^-$ from the beginning of the simulation to this time.}
\label{comparefirsts}
\end{figure}

During the first oscillation, we expect low frequency modes of the $A^-$ mode to be excited. This should be extremely efficient up to a maximum wavenumber, $(k/ a) \approx m \phi_0( {\alpha }/f)$.  In each simulation we have a slightly different value of the inflaton field at the end of inflation, $\phi_0$, depending on the value of the coupling $\alpha / f$ -- although they vary only by only about 15\%, as shown in Table~\ref{tab:phi0}. Fig.~\ref{comparefirsts} is a comparison of the strength of this tachyonic instability including early effects of backreaction and rescattering (which seems to be largely missing from analytic studies of preheating in these models) showing the effect of this first tachyonic regime on the power in the gauge field.

\begin{figure}[h!] 
\centering
\includegraphics[width=0.45\columnwidth]{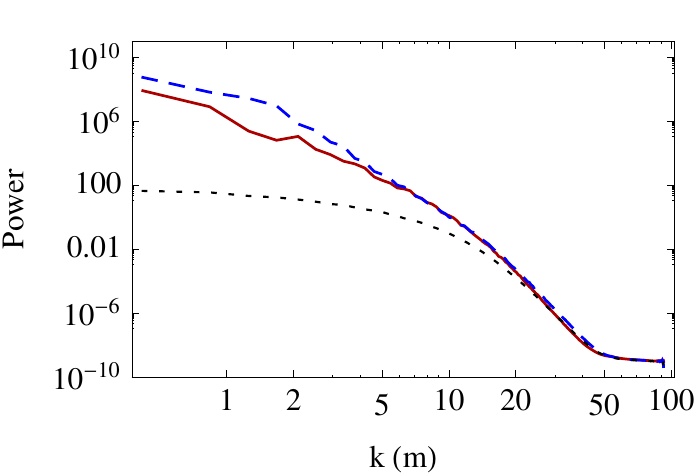}\includegraphics[width=0.45\columnwidth]{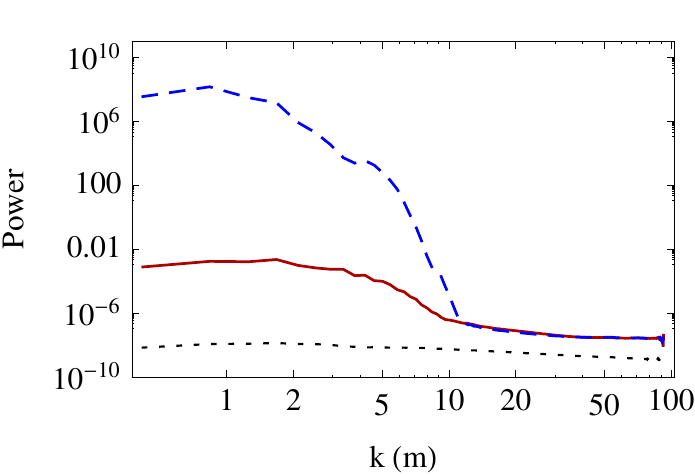}
\caption{The final power spectra of $A^+$ (red, solid), $A^-$ (blue, dashed) and $\phi$ (black, dotted) for a simulation in which the coupling is $\alpha/f = 60\,m_{\rm pl}^{-1}$.  Even though the tachyonic process is only efficient for one of the polarizations, the $A^-$, of the gauge field, there is significant back-reaction onto the modes of $\phi$ which source modes of $A^+$.  The left panel shows the final spectrum for a simulation of our model.  The right panel shows the same simulation, however, with the interaction term eliminated from the equations of motion for the axion field.  This comparison shows that a lot of power is transferred between polarizations, mediated by the axion. \label{final state}  Both panels are evaluated at $a=8.$}
\label{finalstate}
\end{figure}

For moderate couplings, $\alpha /f  \gtrsim 50\,m_{\rm pl}^{-1}$ we see that this first wave of tachyonic instability is extremely efficient and an $\mathcal{O}(1)$ fraction of the total energy density of the Universe is already deposited into the gauge fields.  This is a highly-tachyonic regime.  By the time that the homogeneous mode of the axion has completed its second zero crossing, it is already incoherent and the majority of the energy of the axion is in higher-momentum modes.  Fig.~\ref{finalstate} shows the final spectra of the gauge field for one of these cases.  Note that the spectra do not exhibit any discernible signs of band structure. This is expected from the analysis of Section~\ref{sec:analytics}. During the first tachyonic regime all modes within the tachyonic window get amplified in the first oscillation. The band structures form due to resonances after multiple oscillations. However, in these cases, multiple coherent oscillations of the axion background are prevented  due to the strong gauge-field back reaction. This back reaction is so strong that  at the largest couplings probed here the homogeneous axion condensate does not cross zero. We call this period the early tachyonic regime; the period of strong tachyonic growth that occurs during the first oscillation of the axion.

\begin{figure}[h!] 
\centering
\includegraphics[width=0.45\columnwidth]{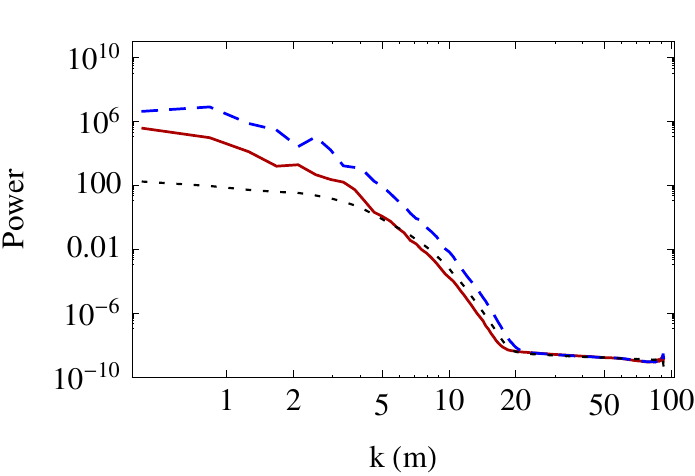}\includegraphics[width=0.45\columnwidth]{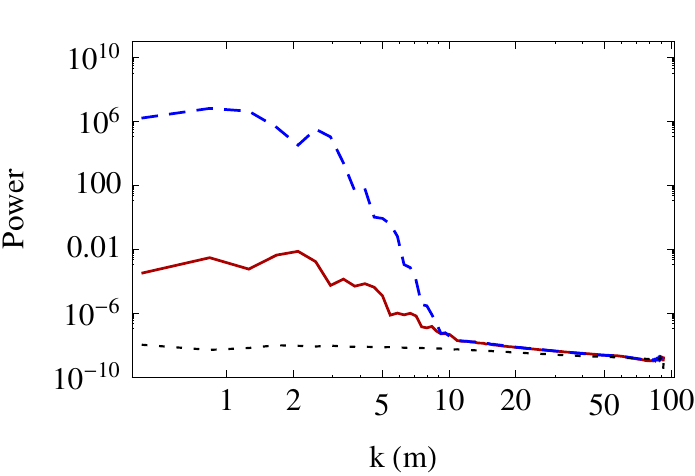}
\caption{The final power spectra of $A^+$ (red, solid), $A^-$ (blue, dashed) and $\phi$ (black, dotted) for a simulation in which the coupling is $\alpha/f = 45\,m_{\rm pl}^{-1}$.  In this case, the final spectrum is not polarized; the power in each of the two modes is almost identical.  The left panel shows the final spectrum for a simulation of our model.  The right panel shows the same simulation, however, with the interaction term eliminated from the equations of motion for the axion field.  This comparison shows that the transfer of power between modes is still important in this regime. Both panels are evaluated at $a=8$.}
\label{finalstate2}
\end{figure}

Naively, one might have predicted that these larger couplings would have lead to a preheated state that is highly polarized.  The lack of oscillations of the axion effectively prevents the tachyonic regime of the $A^+$ mode from developing. On the other hand, the $A^-$ mode is strongly produced as the axion condensate relaxes to zero. Further, once the axion condensate has become unimportant, any energy transfer essentially ceases.  While this is certainly the case for the process of interest, during preheating, the excited $A^-$ modes strongly re-scatter off $\phi$ and source $A^+$  modes. This process is very efficient on sub-horizon scales, and the resulting spectra show very little difference in power between each polarization on these scales. This effect is demonstrated in Fig.~\ref{finalstate}, where we plot the final spectra of the gauge modes, $A^\pm$, and the axion, $\phi$, in the full model. We also show the effect of artificially switching-off this rescattering by eliminating the terms in the equation of motion for the axion that couple it to the gauge field, the term proportional to $\alpha/f$ in Eq.~\ref{axioneom}. This prevents non-linearities from developing in the axion and the subsequent rescattering of the gauge modes.  Note that the final state in these cases is strongly polarized, which demonstrates the efficiency of re-scattering.  In the case where we artificially block this back-reaction, our simulations do not conserve energy and the Hubble parameter rises during the tachyonic phases; this causes the spectra for the gauge fields to be larger than expected in the full simulations by a small factor.  The right panel of Fig.~\ref{finalstate} (as well as Figs.~\ref{finalstate2} and \ref{finalstate3}) should be considered illustrative. Comparing this to Fig.\ \ref{fig:WKB_results} we see that in the absence of back-reaction the amplitudes of two polarizations differ by several orders of magnitude, as expected by our semi-analytic study. 

Even in the presence of back-reaction, the spectra are still polarized in the infrared (the spectra here differ by a couple orders of magnitude, see left panel of Fig.~\ref{finalstate}), suggesting that  the asymmetric  long-wavelength modes might have some observational consequence.

When the tachyonic growth factor is very large (Fig.~\ref{fig:WKB_results}), the energy transferred to the gauge field ($A^-$) after the first oscillation is comparable to the initial energy stored in the inflaton condensate, and the Universe preheats almost instantaneously. The couplings that lead to instantaneous preheating can be estimated from the results of Section \ref{sec:analytics}. The energy density of the background inflaton is
\begin{equation}
\rho_\phi = {1\over 2} \left ( {\partial_t \phi } \right )^2 + {1\over 2} m^2 \phi^2 \approx 0.05\, m^2 m_{\rm Pl}^2 \,
\end{equation}
where we used the value of $\phi$ at the end of the first tachyonic stage. The energy density of the gauge field can be calculated from Eq.\ \ref{eq:rhoEM} to be
\begin{equation}
\rho_{EM} = {1\over a^4} \int d^3 k \left ( |\partial_\tau A^- |^2 + k^2  | A^- |^2  \right ) \approx   {1\over  2\pi^2 a^4} \int dk k k^3  e^{2X_k} \approx  m^4 {k^4\over 4 m^4}{1\over  2\pi^2 a^4} e^{X_k^{\rm max}}
\end{equation}
where we dropped the time-derivative term, since the mode functions do not grow between tachyonic regions. Further we only considered the $A^-$ mode, since during the first tachyonic region the $A^+$ mode is exponentially smaller, hence it can be safely ignored in the calculation of the energy density. The last approximate equality is based on the fact that the function $e^{X_k}$ is sharply peaked. Using the values given in Fig.\ \ref{fig:WKB_results}, we get $\rho_{\rm EM} / \rho_\phi \sim 10^{-4}, 10^{-2}, 10^2$ for $m_{\rm Pl} \alpha/f = 35,45,55$ respectively after the first tachyonic regime at $a(t)\approx 2$. The first two values agree with Fig.\ \ref{emendate}. The analytical result for $m_{\rm Pl} \alpha/f = 55$ signifies that we have entered the region of strong energy transfer and strong back-reaction, hence we cannot treat the axion field as decoupled. These estimations immediately show the region of couplings, for which instantaneous preheating can occur, where full non-linear simulations are unavoidable.

There are additionally cases, however, in which the tachyonic instabilities are present, but not sufficiently efficient to deposit an $\mathcal{O}(1)$ fraction of the total energy into the gauge field during the first oscillation. In these cases, it can take up to ten oscillations of the homogeneous mode of $\phi$ before the Universe is preheated.  It is still possible for the Universe to totally preheat in these cases, it just takes longer.  This regime also results in an unpolarized final state, as seen in Fig.~\ref{finalstate2}, where the axion, again, plays a significant role in balancing the two helicity states.

%%%%%%%%%%%%%%%%%%%%%%%%%%%%%%%%%%%%%%%
%%%%%%%%%%%%%%%%%%%%%%%%%%%%%%%%%%%%%%%

\subsection{Parametric resonance}

For lower values of the coupling, $\alpha/f$, the efficiency of the early tachyonic regime is not high enough to completely preheat the Universe.  At the end of inflation, in the homogeneous field limit, the potential and kinetic energy of the inflaton are approximately equal and so
\begin{equation}
H\approx  \sqrt{\frac{8\pi}{3}}\frac{\dot{\phi}}{m_{\rm pl}}  \approx 0.4\,m,
\end{equation}
which defines the smallest wavelength that is dynamical.  Since the maximum wavenumber that can be amplified is
\begin{equation}
\frac{k_{\rm max}}{ a(t)} =  \frac{\alpha}{f} m \phi_0,
\end{equation}
we see that the band that gets amplified shrinks as $\alpha/f$ gets smaller. Examining Fig.~\ref{fig:WKB_results}, we see that both the growth factor, as well as the regime of amplified wavenumbers, shrink as the coupling $\alpha/f$ gets smaller and the early tachyonic regime is not sufficient to transfer most of the inflaton energy into the gauge fields. 

\begin{figure}[h!] 
\centering
\includegraphics[width=0.45\columnwidth]{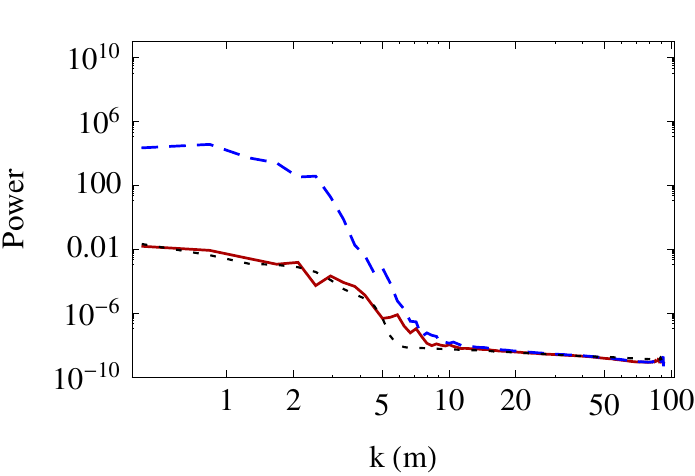}\includegraphics[width=0.45\columnwidth]{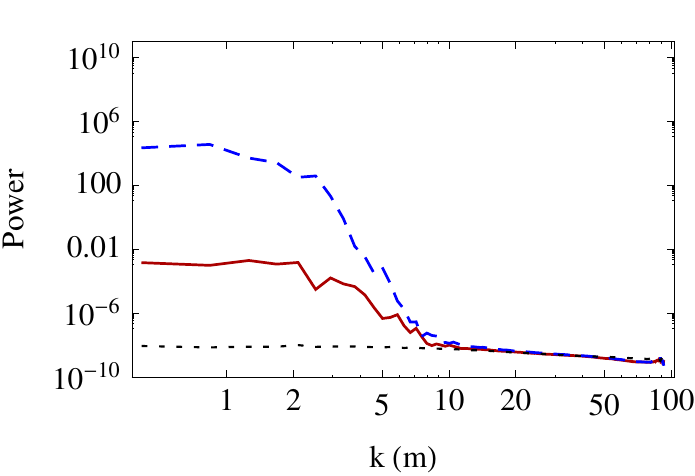}\caption{The final power spectra of $A^+$ (red, solid), $A^-$ (blue, dashed) and $\phi$ (black, dotted) for a simulation in which the coupling is $\alpha/f = 35\,m_{\rm pl}^{-1}$.  In this case, we see a more complex spectrum whose modes have been amplified by a number of (varying) resonance bands.  The left panel shows the final spectrum for a simulation of our model.  The right panel shows the same simulation, however, with the interaction term eliminated from the equations of motion for the axion field.  In both panels the initial asymmetry between the power in the two helicity modes persists until the end of the simulation.  Both panels are evaluated at $a=8$.}
\label{finalstate3}
\end{figure}

While at low couplings, gauge-field production during the early tachyonic-regime is not  strong enouigh to reheat the Universe, the persistence of coherent oscillations of the axion condensate allows for parametric resonance. Parametric resonance continues to deposit power into the gauge field -- independently of polarization-- for many oscillations.  In Fig.~\ref{finalstate3} we show the power spectra of the gauge field polarizations when $\alpha/f = 30\,m_{\rm pl}^{-1}$. Since many different modes have passed through different parametric-resonance bands over many  oscillations, a complex spectral structure forms. However, this spectral structure is shared between the two polarizations equally, because in the limit of many oscillations the two polarizations obey identical equations. While the final state is polarized, this is a residual effect of the initial conditions.  The formation of these spectral bands can be explained in the WKB approximation  (for $q\gg 1$) by a process analogous to multiple scatterings from a periodic potential leading to constructive and destructive interference (as discussed in \cite{Dufaux:2006ee}), as well as through Floquet theory (for $q\lesssim 1$).

However, the question that arises by inspecting Fig.\ \ref{emendate} is the inability of lower couplings ($\alpha/f \lesssim 40 m_{\rm Pl}^{-1}$) to fully reheat the Universe. The parameters of the Mathieu equation of Eq.\ \ref{eq:matparams} in an expanding matter-dominated universe become
\begin{equation}
A_k = 4\( {k\over m\, a(t) } \) ^2 + {2\over 9 t^2}, \quad
q = \mp 2  {k\over m\, a(t)} \, \frac{\alpha}{f} {\phi_0 \over t} \, ,
\end{equation}
where we took the scale factor to evolve as $a(t)\sim t^{2/3}$ and the envelope of the inflaton oscillations to decay as $\phi_0(t) = \phi_0 \, a^{-3/2}= \phi_0 \, t^{-1}$. We thus see that as time progresses the product of the coupling with the axion amplitude becomes smaller. Since this product controls the strength of the parametric resonance, we can call it the ``effective coupling''. For $\alpha /f = 40m_{\rm Pl}^{-1}$, the effective coupling becomes unity at $t\approx  9m$ or equivalently $a(t) \approx 4$. The analysis of \cite{ArmendarizPicon:2007iv} shows that a pseudo-scalar inflaton derivatively coupled to $U(1)$ gauge fields with ${\cal O}(1)$ coupling does not fully preheat, since the growth rate is similar in strength to the red-shifting of the gauge field amplitude due to the expansion of the Universe. Transferring this known result to our case simply means that, in models where most of the inflaton energy has not been transferred to the gauge fields by the time the effective coupling equals unity, no significant preheating occurs. This both explains the behavior seen in Fig.\ \ref{emendate} as well as further demonstrates the importance of the first few axion oscillations for preheating into gauge fields. In any case, perturbative decay of the axion to gauge field is still present and can eventually reheat the Universe, as discussed in Section \ref{sec:perturbative}.
 
\subsection{Monodromy potential}

\begin{figure}[h!] 
\centering
\includegraphics[width=0.45\columnwidth]{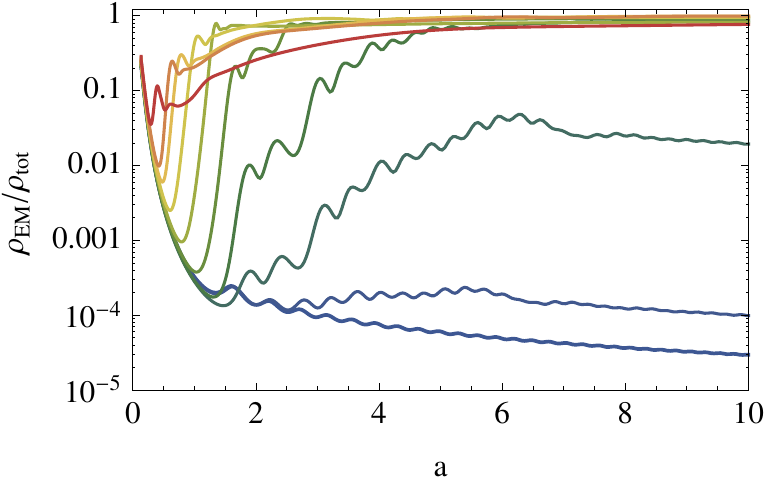}
\includegraphics[width=0.43\columnwidth]{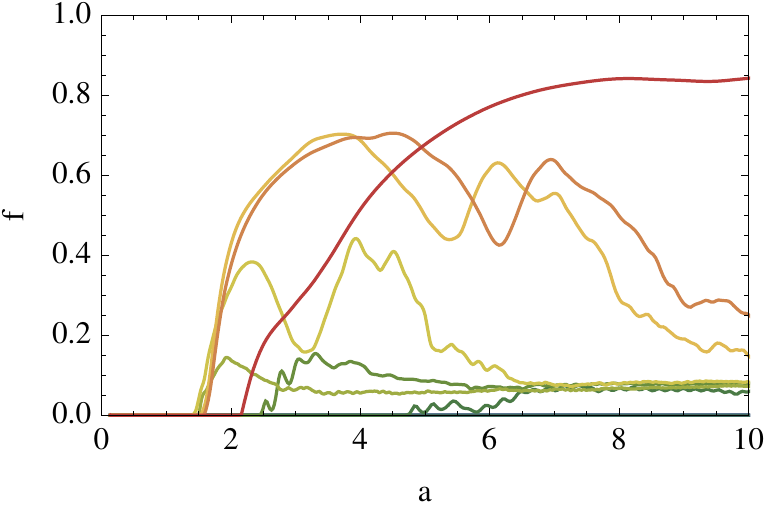}
\caption{The left panel shows the total energy fraction in the gauge field $\rho_{\rm em}/\rho_{\rm total}$ as a function of time for a variety of couplings. These range from $\alpha/f = 120\,m_{\rm pl}^{-1}$ to $\alpha/f = 30\,m_{\rm pl}^{-1}$ in increments of $\alpha/f = 10\,m_{\rm pl}^{-1}$. The couplings go 
from largest on the top (red) to the lowest on the bottom (blue) along the rainbow spectrum. The right panel shows the fraction of the axion energy contained in regions where $\rho_\phi> 4\left<\rho_\phi\right>$ (defined by Eq.~\ref{defoff}) using the same color scheme. This is a proxy to test for the relative abundance of oscillons in these simulations.  
\label{emmonodrom}}
\end{figure}

It is worth verifying that the results presented here are (at least somewhat) insensitive to the specific form of our axion potential.  To test this, we study reheating following monodromy inflation in the potential of Eq.~\ref{eqn:monopot}.  This potential introduces another scale to the problem and causes the oscillations to be increasingly anharmonic when the field probes the region $\phi\gtrsim \phi_c$. As noted above, the fluctuations observed in the CMB are insensitive to the value of $\phi_c$ (as in Section~\ref{background}) and thus we can treat it as a free parameter. For this section we choose $\phi_c = 0.02\, m_{\rm pl}$.  We then chose a set of couplings, $ \alpha / f$, and checked to see how much energy is transferred into the coupled gauge field. Fig.~\ref{emmonodrom} shows the effect on the reheating efficiency for this potential.  

As can be seen from Fig.~\ref{emmonodrom}, monodromy inflation is less efficient at reheating into gauge fields than quadratic, $m^2\phi^2$, inflation, typically requiring larger couplings to achieve reheating within the first axion oscillation.  This is not particularly surprising, as the axion typically rolls at a slower rate relative to the Hubble rate on this potential, and consequently gauge-field production is lower. This is evidenced by the larger allowed values of the coupling $\alpha/f$ by the black-hole abundance bounds, which are  allowed to be around approximately 50\% larger than the $m^2\phi^2$ couplings. We find similar behavior in the reheating epoch, the lowest coupling that completely reheats in this case is  $\alpha/f \sim  70 \,m_{\rm pl}^{-1}$ which should be compared with $\alpha/f \sim  45 \,m_{\rm pl}^{-1}$ for $m^2\phi^2$. This result is due to the fact that the efficiency of the tachyonic regime is sensitive to the axion velocity.  

It is known that under certain conditions pseudo-stable inflaton lumps, or oscillons \cite{Amin:2010dc, Amin:2010xe, Amin:2011hj, Lozanov:2014zfa}, can form from the post-inflationary detritus. The interplay between scalar and gauge fields in oscillons has only been studied for models where the scalar field is charged and the gauge field is non-Abelian \cite{Sfakianakis:2012bq, Graham:2007ds}. The oscillons that form from the fragmentation of the inflaton following inflation can be long lived and have drastic consequences for theories of the early Universe. This is because they behave like pressure-less dust and can change the early expansion history.  Our methodology is sufficiently robust to allow us to study the creation and decay of these structures, as well as to probe the effect of these structures on our reheating history.  We test for the formation of oscillons in our simulations, by computing the fraction of energy contained in regions of high density compared to the average background energy density.  Specifically, we compute the total energy in regions where the local density is greater than four times the average density, and then compare this to the total energy in the box \cite{Amin:2011hj},
\begin{equation}
\label{defoff}
f = \frac{\int_{\rho_\phi>4\left<\rho_\phi\right>} \rho_\phi\,dV}{\int \rho_\phi \,dV}.
\end{equation}
We note that our definition differs from that in \cite{Amin:2011hj}, since we only calculate the energy in the axion field, $\rho_\phi$, when we compute the numerator and denominator and we restrict the integral in the numerator to regions of very high energy density.  We also only integrate the numerator for regions where the axion energy density is four times the average axion energy density (instead of twice the axion energy density).  The right hand panel of Fig.~\ref{emmonodrom} shows this statistic and shows that the time at which oscillons form is highly dependent on the coupling, suggesting that the gauge fields play a role in the creation (and possible decay) of these structures.  We delay a full treatment of oscillons for a future publication.  
%%%%%%%%%%%%%%%%%%%%%%%%%%%%%%%%%%%%%%%
%%%%%%%%%%%%%%%%%%%%%%%%%%%%%%%%%%%%%%%
%%%%%%%%%%%%%%%%%%%%%%%%%%%%%%%%%%%%%%%

\section{Conclusions}

In this work we have studied preheating into $U(1)$ gauge fields following axion inflation. The shift-symmetry of the axion means that the form of the couplings to matter fields is highly restricted. A class of allowed couplings is via derivative interactions, and consequently dimension-five couplings to gauge fields via Chern-Simons terms are expected in the theory from an effective field theory point of view. The size of such a coupling is unknown, however, the effect of rescattering of helically polarized gauge bosons off the axion condensate  and the subsequent generation of curvature fluctuations during the inflationary epoch places an upper bound. In this work we have explored the phenomenological consequences of such a coupling on the post-inflationary evolution of axion-driven inflationary models.

Using lattice simulations and semi-analytic methods we have shown that preheating into Abelian gauge fields via these Chern-Simons interactions can be extremely efficient. In particular, at the middle to upper range of the couplings allowed by constraints due to over-production of primordial black holes, we find that reheating is essentially instantaneous, proceeding via a phase of early tachyonic resonance and completing within a single oscillation of the axion. The resulting Universe ends up in an un-polarized state due to strong rescattering effects on scales that are sub-horizon during reheating. Scattering of amplified gauge field modes into axion fluctuations generates the second polarization extremely efficiently. On super-horizon scales, the asymmetry that develops due to the tachyonic instability of one of the gauge modes during inflation, remains. The Universe in these high coupling cases is radiation dominated and is characterized by very efficient preheating where non-linear dynamics and back reaction become important almost immediately.  As the coupling is decreased, this phase of early tachyonic resonance is weakened, and the axion oscillates multiple times before preheating completes. During these multiple oscillations equal levels of both polarizations of the gauge field are excited due to tachyonic and parametric resonance. Decreasing the coupling further yields a brief window where parametric-resonance effects become important, before preheating abruptly shuts off and non-linear effects cease to be important. At these lower couplings, non-linear effects are never important and the Universe reheats perturbatively due to the decay of the axion into gauge bosons with a reheating temperature near $10^9$ GeV. Note that all of the couplings we have considered in this work are far below those values that give any observable effects in the CMB. At the couplings that saturate the bounds considered in the recent Planck paper \cite{Ade:2015lrj}, reheating would be essentially instantaneous and, perhaps, accompanied by an overproduction of primordial black holes.

We studied reheating in two different axion potentials: the simplest model of chaotic inflation with a monomial potential, $m^2\phi^2$, and the simplest model of axion monodromy inflation. Phenomenologically, these two potentials have similar (p)reheating behavior. However, generically for a given coupling, the monodromy potential has a lower reheating efficiency compared to the chaotic inflation case due to the slower initial axion velocity.  The anharmonic nature of the axion increases the features of preheating in this model and warrants further study.

For the case of monodromy inflation, it is well known that at the end of inflation pseudo-stable classical lumps of the axion field -- oscillons -- can form and lead to a period of matter domination before the onset of reheating. Our numerical investigations suggest that these couplings to gauge fields only strengthen the formation of these oscillons and leads to  their formation at an earlier epoch compared to the uncoupled case. However, the Universe remains radiation dominated due to the bath of gauge bosons produced. We leave a full study of oscillons in these models to future work.

%%%%%%%%%%%%%%%%%%%%%%%%%%%%%%%%%%%%%%%
%%%%%%%%%%%%%%%%%%%%%%%%%%%%%%%%%%%%%%%
%%%%%%%%%%%%%%%%%%%%%%%%%%%%%%%%%%%%%%%

\acknowledgments
We thank Mustafa Amin for useful discussions. PA gratefully acknowledges support from a Starting Grant of the European Research Council (ERC STG grant 279617) and the hospitality of the Department of Applied Mathematics and Theoretical Physics at the University of Cambridge where some of this work was completed.  JTG and TRS  are supported by the National Science Foundation, PHY-1414479.  We acknowledge the  National Science Foundation, the Research Corporation for Science Advancement and the Kenyon College Department of Physics for providing the hardware used to carry out these simulations. EIS gratefully acknowledges support from a Fortner Fellowship at the University of Illinois at Urbana-Champaign.

\bibliographystyle{JHEP}
\bibliography{AxReheat}

\providecommand{\href}[2]{#2}\begingroup\raggedright\begin{thebibliography}{100}

\bibitem{Ade:2014xna}
{\bf BICEP2 Collaboration} Collaboration, P.~Ade et~al., {\it {Detection of
  B-Mode Polarization at Degree Angular Scales by BICEP2}},  {\em
  Phys.Rev.Lett.} {\bf 112} (2014) 241101,
  [\href{http://arxiv.org/abs/1403.3985}{{\tt arXiv:1403.3985}}].

\bibitem{Adam:2014bub}
{\bf Planck Collaboration} Collaboration, R.~Adam et~al., {\it {Planck
  intermediate results. XXX. The angular power spectrum of polarized dust
  emission at intermediate and high Galactic latitudes}},
  \href{http://arxiv.org/abs/1409.5738}{{\tt arXiv:1409.5738}}.

\bibitem{Ade:2015tva}
{\bf BICEP2 Collaboration, Planck Collaboration} Collaboration, P.~Ade et~al.,
  {\it {A Joint Analysis of BICEP2/Keck Array and Planck Data}},  {\em
  Phys.Rev.Lett.} (2015) [\href{http://arxiv.org/abs/1502.00612}{{\tt
  arXiv:1502.00612}}].

\bibitem{Mukhanov:1981xt}
V.~F. Mukhanov and G.~V. Chibisov, {\it {Quantum Fluctuation and Nonsingular
  Universe. (In Russian)}},  {\em JETP Lett.} {\bf 33} (1981) 532--535.

\bibitem{Lyth:1996im}
D.~H. Lyth, {\it {What would we learn by detecting a gravitational wave signal
  in the cosmic microwave background anisotropy?}},  {\em Phys.Rev.Lett.} {\bf
  78} (1997) 1861--1863, [\href{http://arxiv.org/abs/hep-ph/9606387}{{\tt
  hep-ph/9606387}}].

\bibitem{Freese:1990rb}
K.~Freese, J.~A. Frieman, and A.~V. Olinto, {\it {Natural inflation with pseudo
  - Nambu-Goldstone bosons}},  {\em Phys.Rev.Lett.} {\bf 65} (1990) 3233--3236.

\bibitem{Adams:1992bn}
F.~C. Adams, J.~R. Bond, K.~Freese, J.~A. Frieman, and A.~V. Olinto, {\it
  {Natural inflation: Particle physics models, power law spectra for large
  scale structure, and constraints from COBE}},  {\em Phys. Rev.} {\bf D47}
  (1993) 426--455, [\href{http://arxiv.org/abs/hep-ph/9207245}{{\tt
  hep-ph/9207245}}].

\bibitem{Banks:2003sx}
T.~Banks, M.~Dine, P.~J. Fox, and E.~Gorbatov, {\it {On the possibility of
  large axion decay constants}},  {\em JCAP} {\bf 0306} (2003) 001,
  [\href{http://arxiv.org/abs/hep-th/0303252}{{\tt hep-th/0303252}}].

\bibitem{Dimopoulos:2005ac}
S.~Dimopoulos, S.~Kachru, J.~McGreevy, and J.~G. Wacker, {\it {N-flation}},
  {\em JCAP} {\bf 0808} (2008) 003,
  [\href{http://arxiv.org/abs/hep-th/0507205}{{\tt hep-th/0507205}}].

\bibitem{Easther:2005zr}
R.~Easther and L.~McAllister, {\it {Random matrices and the spectrum of
  N-flation}},  {\em JCAP} {\bf 0605} (2006) 018,
  [\href{http://arxiv.org/abs/hep-th/0512102}{{\tt hep-th/0512102}}].

\bibitem{Bachlechner:2014hsa}
T.~C. Bachlechner, M.~Dias, J.~Frazer, and L.~McAllister, {\it {A New Angle on
  Chaotic Inflation}},  \href{http://arxiv.org/abs/1404.7496}{{\tt
  arXiv:1404.7496}}.

\bibitem{Liddle:1998jc}
A.~R. Liddle, A.~Mazumdar, and F.~E. Schunck, {\it {Assisted inflation}},  {\em
  Phys.Rev.} {\bf D58} (1998) 061301,
  [\href{http://arxiv.org/abs/astro-ph/9804177}{{\tt astro-ph/9804177}}].

\bibitem{Kim:2004rp}
J.~E. Kim, H.~P. Nilles, and M.~Peloso, {\it {Completing natural inflation}},
  {\em JCAP} {\bf 0501} (2005) 005,
  [\href{http://arxiv.org/abs/hep-ph/0409138}{{\tt hep-ph/0409138}}].

\bibitem{Long:2014dta}
C.~Long, L.~McAllister, and P.~McGuirk, {\it {Aligned Natural Inflation in
  String Theory}},  {\em Phys.Rev.} {\bf D90} (2014) 023501,
  [\href{http://arxiv.org/abs/1404.7852}{{\tt arXiv:1404.7852}}].

\bibitem{Burgess:2014oma}
C.~Burgess and D.~Roest, {\it {Inflation by Alignment}},
  \href{http://arxiv.org/abs/1412.1614}{{\tt arXiv:1412.1614}}.

\bibitem{McAllister:2008hb}
L.~McAllister, E.~Silverstein, and A.~Westphal, {\it {Gravity Waves and Linear
  Inflation from Axion Monodromy}},  {\em Phys.Rev.} {\bf D82} (2010) 046003,
  [\href{http://arxiv.org/abs/0808.0706}{{\tt arXiv:0808.0706}}].

\bibitem{Silverstein:2008sg}
E.~Silverstein and A.~Westphal, {\it {Monodromy in the CMB: Gravity Waves and
  String Inflation}},  {\em Phys.Rev.} {\bf D78} (2008) 106003,
  [\href{http://arxiv.org/abs/0803.3085}{{\tt arXiv:0803.3085}}].

\bibitem{McAllister:2014mpa}
L.~McAllister, E.~Silverstein, A.~Westphal, and T.~Wrase, {\it {The Powers of
  Monodromy}},  {\em JHEP} {\bf 1409} (2014) 123,
  [\href{http://arxiv.org/abs/1405.3652}{{\tt arXiv:1405.3652}}].

\bibitem{Kaloper:2011jz}
N.~Kaloper, A.~Lawrence, and L.~Sorbo, {\it {An Ignoble Approach to Large Field
  Inflation}},  {\em JCAP} {\bf 1103} (2011) 023,
  [\href{http://arxiv.org/abs/1101.0026}{{\tt arXiv:1101.0026}}].

\bibitem{Marchesano:2014mla}
F.~Marchesano, G.~Shiu, and A.~M. Uranga, {\it {F-term Axion Monodromy
  Inflation}},  {\em JHEP} {\bf 1409} (2014) 184,
  [\href{http://arxiv.org/abs/1404.3040}{{\tt arXiv:1404.3040}}].

\bibitem{Blumenhagen:2014gta}
R.~Blumenhagen and E.~Plauschinn, {\it {Towards Universal Axion Inflation and
  Reheating in String Theory}},  {\em Phys.Lett.} {\bf B736} (2014) 482--487,
  [\href{http://arxiv.org/abs/1404.3542}{{\tt arXiv:1404.3542}}].

\bibitem{Hebecker:2014eua}
A.~Hebecker, S.~C. Kraus, and L.~T. Witkowski, {\it {D7-Brane Chaotic
  Inflation}},  {\em Phys.Lett.} {\bf B737} (2014) 16--22,
  [\href{http://arxiv.org/abs/1404.3711}{{\tt arXiv:1404.3711}}].

\bibitem{Cai:2014vua}
Y.-F. Cai, F.~Chen, E.~G.~M. Ferreira, and J.~Quintin, {\it {A new model of
  axion monodromy inflation and its cosmological implications}},
  \href{http://arxiv.org/abs/1412.4298}{{\tt arXiv:1412.4298}}.

\bibitem{Pajer:2013fsa}
E.~Pajer and M.~Peloso, {\it {A review of Axion Inflation in the era of
  Planck}},  {\em Class.Quant.Grav.} {\bf 30} (2013) 214002,
  [\href{http://arxiv.org/abs/1305.3557}{{\tt arXiv:1305.3557}}].

\bibitem{Baumann:2014nda}
D.~Baumann and L.~McAllister, {\it {Inflation and String Theory}},
  \href{http://arxiv.org/abs/1404.2601}{{\tt arXiv:1404.2601}}.

\bibitem{Amin:2014eta}
M.~A. Amin, M.~P. Hertzberg, D.~I. Kaiser, and J.~Karouby, {\it
  {Nonperturbative Dynamics Of Reheating After Inflation: A Review}},
  \href{http://arxiv.org/abs/1410.3808}{{\tt arXiv:1410.3808}}.

\bibitem{Carroll:1991zs}
S.~M. Carroll and G.~B. Field, {\it {The Einstein equivalence principle and the
  polarization of radio galaxies}},  {\em Phys.Rev.} {\bf D43} (1991) 3789.

\bibitem{Garretson:1992vt}
W.~D. Garretson, G.~B. Field, and S.~M. Carroll, {\it {Primordial magnetic
  fields from pseudoGoldstone bosons}},  {\em Phys.Rev.} {\bf D46} (1992)
  5346--5351, [\href{http://arxiv.org/abs/hep-ph/9209238}{{\tt
  hep-ph/9209238}}].

\bibitem{Prokopec:2001nc}
T.~Prokopec, {\it {Cosmological magnetic fields from photon coupling to
  fermions and bosons in inflation}},
  \href{http://arxiv.org/abs/astro-ph/0106247}{{\tt astro-ph/0106247}}.

\bibitem{Anber:2009ua}
M.~M. Anber and L.~Sorbo, {\it {Naturally inflating on steep potentials through
  electromagnetic dissipation}},  {\em Phys.Rev.} {\bf D81} (2010) 043534,
  [\href{http://arxiv.org/abs/0908.4089}{{\tt arXiv:0908.4089}}].

\bibitem{Barnaby:2011vw}
N.~Barnaby, R.~Namba, and M.~Peloso, {\it {Phenomenology of a Pseudo-Scalar
  Inflaton: Naturally Large Nongaussianity}},  {\em JCAP} {\bf 1104} (2011)
  009, [\href{http://arxiv.org/abs/1102.4333}{{\tt arXiv:1102.4333}}].

\bibitem{Barnaby:2011qe}
N.~Barnaby, E.~Pajer, and M.~Peloso, {\it {Gauge Field Production in Axion
  Inflation: Consequences for Monodromy, non-Gaussianity in the CMB, and
  Gravitational Waves at Interferometers}},  {\em Phys.Rev.} {\bf D85} (2012)
  023525, [\href{http://arxiv.org/abs/1110.3327}{{\tt arXiv:1110.3327}}].

\bibitem{Ferreira:2014zia}
R.~Z. Ferreira and M.~S. Sloth, {\it {Universal Constraints on Axions from
  Inflation}},  {\em JHEP} {\bf 1412} (2014) 139,
  [\href{http://arxiv.org/abs/1409.5799}{{\tt arXiv:1409.5799}}].

\bibitem{Adshead:2013qp}
P.~Adshead, E.~Martinec, and M.~Wyman, {\it {Gauge fields and inflation: Chiral
  gravitational waves, fluctuations, and the Lyth bound}},  {\em Phys. Rev.}
  {\bf D88} (2013), no.~2 021302, [\href{http://arxiv.org/abs/1301.2598}{{\tt
  arXiv:1301.2598}}].

\bibitem{Shiraishi:2013kxa}
M.~Shiraishi, A.~Ricciardone, and S.~Saga, {\it {Parity violation in the CMB
  bispectrum by a rolling pseudoscalar}},  {\em JCAP} {\bf 1311} (2013) 051,
  [\href{http://arxiv.org/abs/1308.6769}{{\tt arXiv:1308.6769}}].

\bibitem{Cook:2013xea}
J.~L. Cook and L.~Sorbo, {\it {An inflationary model with small scalar and
  large tensor nongaussianities}},  {\em JCAP} {\bf 1311} (2013) 047,
  [\href{http://arxiv.org/abs/1307.7077}{{\tt arXiv:1307.7077}}].

\bibitem{Brandenberger:2008kn}
R.~H. Brandenberger, A.~Knauf, and L.~C. Lorenz, {\it {Reheating in a Brane
  Monodromy Inflation Model}},  {\em JHEP} {\bf 0810} (2008) 110,
  [\href{http://arxiv.org/abs/0808.3936}{{\tt arXiv:0808.3936}}].

\bibitem{ArmendarizPicon:2007iv}
C.~Armendariz-Picon, M.~Trodden, and E.~J. West, {\it {Preheating in
  derivatively-coupled inflation models}},  {\em JCAP} {\bf 0804} (2008) 036,
  [\href{http://arxiv.org/abs/0707.2177}{{\tt arXiv:0707.2177}}].

\bibitem{Braden:2010wd}
J.~Braden, L.~Kofman, and N.~Barnaby, {\it {Reheating the Universe After
  Multi-Field Inflation}},  {\em JCAP} {\bf 1007} (2010) 016,
  [\href{http://arxiv.org/abs/1005.2196}{{\tt arXiv:1005.2196}}].

\bibitem{Barnaby:2010vf}
N.~Barnaby and M.~Peloso, {\it {Large Nongaussianity in Axion Inflation}},
  {\em Phys.Rev.Lett.} {\bf 106} (2011) 181301,
  [\href{http://arxiv.org/abs/1011.1500}{{\tt arXiv:1011.1500}}].

\bibitem{Linde:2012bt}
A.~Linde, S.~Mooij, and E.~Pajer, {\it {Gauge field production in supergravity
  inflation: Local non-Gaussianity and primordial black holes}},  {\em
  Phys.Rev.} {\bf D87} (2013), no.~10 103506,
  [\href{http://arxiv.org/abs/1212.1693}{{\tt arXiv:1212.1693}}].

\bibitem{Bugaev:2013fya}
E.~Bugaev and P.~Klimai, {\it {Axion inflation with gauge field production and
  primordial black holes}},  {\em Phys.Rev.} {\bf D90} (2014), no.~10 103501,
  [\href{http://arxiv.org/abs/1312.7435}{{\tt arXiv:1312.7435}}].

\bibitem{Traschen:1990sw}
J.~H. Traschen and R.~H. Brandenberger, {\it {Particle Production During
  Out-of-equilibrium Phase Transitions}},  {\em Phys.Rev.} {\bf D42} (1990)
  2491--2504.

\bibitem{Kofman:1994rk}
L.~Kofman, A.~D. Linde, and A.~A. Starobinsky, {\it {Reheating after
  inflation}},  {\em Phys.Rev.Lett.} {\bf 73} (1994) 3195--3198,
  [\href{http://arxiv.org/abs/hep-th/9405187}{{\tt hep-th/9405187}}].

\bibitem{GarciaBellido:1997wm}
J.~Garcia-Bellido and A.~D. Linde, {\it {Preheating in hybrid inflation}},
  {\em Phys.Rev.} {\bf D57} (1998) 6075--6088,
  [\href{http://arxiv.org/abs/hep-ph/9711360}{{\tt hep-ph/9711360}}].

\bibitem{Khlebnikov:1997di}
S.~Khlebnikov and I.~Tkachev, {\it {Relic gravitational waves produced after
  preheating}},  {\em Phys.Rev.} {\bf D56} (1997) 653--660,
  [\href{http://arxiv.org/abs/hep-ph/9701423}{{\tt hep-ph/9701423}}].

\bibitem{Greene:1997ge}
B.~R. Greene, T.~Prokopec, and T.~G. Roos, {\it {Inflaton decay and heavy
  particle production with negative coupling}},  {\em Phys.Rev.} {\bf D56}
  (1997) 6484--6507, [\href{http://arxiv.org/abs/hep-ph/9705357}{{\tt
  hep-ph/9705357}}].

\bibitem{Parry:1998pn}
M.~Parry and R.~Easther, {\it {Preheating and the Einstein field equations}},
  {\em Phys.Rev.} {\bf D59} (1999) 061301,
  [\href{http://arxiv.org/abs/hep-ph/9809574}{{\tt hep-ph/9809574}}].

\bibitem{Bassett:1998wg}
B.~A. Bassett, D.~I. Kaiser, and R.~Maartens, {\it {General relativistic
  preheating after inflation}},  {\em Phys.Lett.} {\bf B455} (1999) 84--89,
  [\href{http://arxiv.org/abs/hep-ph/9808404}{{\tt hep-ph/9808404}}].

\bibitem{GarciaBellido:1998gm}
J.~Garcia-Bellido, {\it {Preheating the universe in hybrid inflation}},
  \href{http://arxiv.org/abs/hep-ph/9804205}{{\tt hep-ph/9804205}}.

\bibitem{Easther:1999ws}
R.~Easther and M.~Parry, {\it {Gravity, parametric resonance and chaotic
  inflation}},  {\em Phys.Rev.} {\bf D62} (2000) 103503,
  [\href{http://arxiv.org/abs/hep-ph/9910441}{{\tt hep-ph/9910441}}].

\bibitem{Liddle:1999hq}
A.~R. Liddle, D.~H. Lyth, K.~A. Malik, and D.~Wands, {\it {Superhorizon
  perturbations and preheating}},  {\em Phys.Rev.} {\bf D61} (2000) 103509,
  [\href{http://arxiv.org/abs/hep-ph/9912473}{{\tt hep-ph/9912473}}].

\bibitem{Finelli:2001db}
F.~Finelli and S.~Khlebnikov, {\it {Metric perturbations at reheating: The Use
  of spherical symmetry}},  {\em Phys.Rev.} {\bf D65} (2002) 043505,
  [\href{http://arxiv.org/abs/hep-ph/0107143}{{\tt hep-ph/0107143}}].

\bibitem{Bassett:2005xm}
B.~A. Bassett, S.~Tsujikawa, and D.~Wands, {\it {Inflation dynamics and
  reheating}},  {\em Rev.Mod.Phys.} {\bf 78} (2006) 537--589,
  [\href{http://arxiv.org/abs/astro-ph/0507632}{{\tt astro-ph/0507632}}].

\bibitem{Podolsky:2005bw}
D.~I. Podolsky, G.~N. Felder, L.~Kofman, and M.~Peloso, {\it {Equation of state
  and beginning of thermalization after preheating}},  {\em Phys.Rev.} {\bf
  D73} (2006) 023501, [\href{http://arxiv.org/abs/hep-ph/0507096}{{\tt
  hep-ph/0507096}}].

\bibitem{Child:2013ria}
H.~L. Child, J.~Giblin, John~T., R.~H. Ribeiro, and D.~Seery, {\it {Preheating
  with Non-Minimal Kinetic Terms}},  {\em Phys.Rev.Lett.} {\bf 111} (2013)
  051301, [\href{http://arxiv.org/abs/1305.0561}{{\tt arXiv:1305.0561}}].

\bibitem{Rajantie:2000fd}
A.~Rajantie and E.~J. Copeland, {\it {Phase transitions from preheating in
  gauge theories}},  {\em Phys.Rev.Lett.} {\bf 85} (2000) 916,
  [\href{http://arxiv.org/abs/hep-ph/0003025}{{\tt hep-ph/0003025}}].

\bibitem{Copeland:2002ku}
E.~J. Copeland, S.~Pascoli, and A.~Rajantie, {\it {Dynamics of tachyonic
  preheating after hybrid inflation}},  {\em Phys.Rev.} {\bf D65} (2002)
  103517, [\href{http://arxiv.org/abs/hep-ph/0202031}{{\tt hep-ph/0202031}}].

\bibitem{Deskins:2013dwa}
J.~T. Deskins, J.~T. Giblin, and R.~R. Caldwell, {\it {Gauge Field Preheating
  at the End of Inflation}},  {\em Phys.Rev.} {\bf D88} (2013), no.~6 063530,
  [\href{http://arxiv.org/abs/1305.7226}{{\tt arXiv:1305.7226}}].

\bibitem{GarciaBellido:1999sv}
J.~Garcia-Bellido, D.~Y. Grigoriev, A.~Kusenko, and M.~E. Shaposhnikov, {\it
  {Nonequilibrium electroweak baryogenesis from preheating after inflation}},
  {\em Phys.Rev.} {\bf D60} (1999) 123504,
  [\href{http://arxiv.org/abs/hep-ph/9902449}{{\tt hep-ph/9902449}}].

\bibitem{Smit:2002yg}
J.~Smit and A.~Tranberg, {\it {Chern-Simons number asymmetry from CP violation
  at electroweak tachyonic preheating}},  {\em JHEP} {\bf 0212} (2002) 020,
  [\href{http://arxiv.org/abs/hep-ph/0211243}{{\tt hep-ph/0211243}}].

\bibitem{Tranberg:2003gi}
A.~Tranberg and J.~Smit, {\it {Baryon asymmetry from electroweak tachyonic
  preheating}},  {\em JHEP} {\bf 0311} (2003) 016,
  [\href{http://arxiv.org/abs/hep-ph/0310342}{{\tt hep-ph/0310342}}].

\bibitem{Skullerud:2003ki}
J.-I. Skullerud, J.~Smit, and A.~Tranberg, {\it {W and Higgs particle
  distributions during electroweak tachyonic preheating}},  {\em JHEP} {\bf
  0308} (2003) 045, [\href{http://arxiv.org/abs/hep-ph/0307094}{{\tt
  hep-ph/0307094}}].

\bibitem{GarciaBellido:2003wd}
J.~Garcia-Bellido, M.~Garcia-Perez, and A.~Gonzalez-Arroyo, {\it {Chern-Simons
  production during preheating in hybrid inflation models}},  {\em Phys.Rev.}
  {\bf D69} (2004) 023504, [\href{http://arxiv.org/abs/hep-ph/0304285}{{\tt
  hep-ph/0304285}}].

\bibitem{vanTent:2004rc}
B.~van Tent, J.~Smit, and A.~Tranberg, {\it {Electroweak scale inflation,
  inflaton Higgs mixing and the scalar spectral index}},  {\em JCAP} {\bf 0407}
  (2004) 003, [\href{http://arxiv.org/abs/hep-ph/0404128}{{\tt
  hep-ph/0404128}}].

\bibitem{vanderMeulen:2005sp}
M.~van~der Meulen, D.~Sexty, J.~Smit, and A.~Tranberg, {\it {Chern-Simons and
  winding number in a tachyonic electroweak transition}},  {\em JHEP} {\bf
  0602} (2006) 029, [\href{http://arxiv.org/abs/hep-ph/0511080}{{\tt
  hep-ph/0511080}}].

\bibitem{Tranberg:2006dg}
A.~Tranberg, J.~Smit, and M.~Hindmarsh, {\it {Simulations of cold electroweak
  baryogenesis: Finite time quenches}},  {\em JHEP} {\bf 0701} (2007) 034,
  [\href{http://arxiv.org/abs/hep-ph/0610096}{{\tt hep-ph/0610096}}].

\bibitem{Dufaux:2010cf}
J.-F. Dufaux, D.~G. Figueroa, and J.~Garcia-Bellido, {\it {Gravitational Waves
  from Abelian Gauge Fields and Cosmic Strings at Preheating}},  {\em
  Phys.Rev.} {\bf D82} (2010) 083518,
  [\href{http://arxiv.org/abs/1006.0217}{{\tt arXiv:1006.0217}}].

\bibitem{GarciaBellido:2008ab}
J.~Garcia-Bellido, D.~G. Figueroa, and J.~Rubio, {\it {Preheating in the
  Standard Model with the Higgs-Inflaton coupled to gravity}},  {\em Phys.Rev.}
  {\bf D79} (2009) 063531, [\href{http://arxiv.org/abs/0812.4624}{{\tt
  arXiv:0812.4624}}].

\bibitem{DiazGil:2007dy}
A.~Diaz-Gil, J.~Garcia-Bellido, M.~Garcia~Perez, and A.~Gonzalez-Arroyo, {\it
  {Magnetic field production during preheating at the electroweak scale}},
  {\em Phys.Rev.Lett.} {\bf 100} (2008) 241301,
  [\href{http://arxiv.org/abs/0712.4263}{{\tt arXiv:0712.4263}}].

\bibitem{DiazGil:2008tf}
A.~Diaz-Gil, J.~Garcia-Bellido, M.~G. Perez, and A.~Gonzalez-Arroyo, {\it
  {Primordial magnetic fields from preheating at the electroweak scale}},  {\em
  JHEP} {\bf 0807} (2008) 043, [\href{http://arxiv.org/abs/0805.4159}{{\tt
  arXiv:0805.4159}}].

\bibitem{Mazumdar:2008up}
A.~Mazumdar and H.~Stoica, {\it {Exciting gauge field and gravitons in a
  brane-anti-brane annihilation}},  {\em Phys.Rev.Lett.} {\bf 102} (2009)
  091601, [\href{http://arxiv.org/abs/0807.2570}{{\tt arXiv:0807.2570}}].

\bibitem{Allahverdi:2011aj}
R.~Allahverdi, A.~Ferrantelli, J.~Garcia-Bellido, and A.~Mazumdar, {\it
  {Non-perturbative production of matter and rapid thermalization after MSSM
  inflation}},  {\em Phys.Rev.} {\bf D83} (2011) 123507,
  [\href{http://arxiv.org/abs/1103.2123}{{\tt arXiv:1103.2123}}].

\bibitem{Adshead:2012kp}
P.~Adshead and M.~Wyman, {\it {Chromo-Natural Inflation: Natural inflation on a
  steep potential with classical non-Abelian gauge fields}},  {\em Phys. Rev.
  Lett.} {\bf 108} (2012) 261302, [\href{http://arxiv.org/abs/1202.2366}{{\tt
  arXiv:1202.2366}}].

\bibitem{Maleknejad:2012fw}
A.~Maleknejad, M.~M. Sheikh-Jabbari, and J.~Soda, {\it {Gauge Fields and
  Inflation}},  {\em Phys. Rept.} {\bf 528} (2013) 161--261,
  [\href{http://arxiv.org/abs/1212.2921}{{\tt arXiv:1212.2921}}].

\bibitem{Dufaux:2006ee}
J.~F. Dufaux, G.~N. Felder, L.~Kofman, M.~Peloso, and D.~Podolsky, {\it
  {Preheating with trilinear interactions: Tachyonic resonance}},  {\em JCAP}
  {\bf 0607} (2006) 006, [\href{http://arxiv.org/abs/hep-ph/0602144}{{\tt
  hep-ph/0602144}}].

\bibitem{Linde:1981mu}
A.~D. Linde, {\it {A New Inflationary Universe Scenario: A Possible Solution of
  the Horizon, Flatness, Homogeneity, Isotropy and Primordial Monopole
  Problems}},  {\em Phys.Lett.} {\bf B108} (1982) 389--393.

\bibitem{Ade:2013uln}
{\bf Planck Collaboration} Collaboration, P.~Ade et~al., {\it {Planck 2013
  results. XXII. Constraints on inflation}},  {\em Astron.Astrophys.} {\bf 571}
  (2014) A22, [\href{http://arxiv.org/abs/1303.5082}{{\tt arXiv:1303.5082}}].

\bibitem{Peiris:2013opa}
H.~Peiris, R.~Easther, and R.~Flauger, {\it {Constraining Monodromy
  Inflation}},  {\em JCAP} {\bf 1309} (2013) 018,
  [\href{http://arxiv.org/abs/1303.2616}{{\tt arXiv:1303.2616}}].

\bibitem{Easther:2013kla}
R.~Easther and R.~Flauger, {\it {Planck Constraints on Monodromy Inflation}},
  {\em JCAP} {\bf 1402} (2014) 037, [\href{http://arxiv.org/abs/1308.3736}{{\tt
  arXiv:1308.3736}}].

\bibitem{Amin:2011hj}
M.~A. Amin, R.~Easther, H.~Finkel, R.~Flauger, and M.~P. Hertzberg, {\it
  {Oscillons After Inflation}},  {\em Phys.Rev.Lett.} {\bf 108} (2012) 241302,
  [\href{http://arxiv.org/abs/1106.3335}{{\tt arXiv:1106.3335}}].

\bibitem{Ade:2013ydc}
{\bf Planck Collaboration} Collaboration, P.~Ade et~al., {\it {Planck 2013
  Results. XXIV. Constraints on primordial non-Gaussianity}},  {\em
  Astron.Astrophys.} {\bf 571} (2014) A24,
  [\href{http://arxiv.org/abs/1303.5084}{{\tt arXiv:1303.5084}}].

\bibitem{Carr:2009jm}
B.~Carr, K.~Kohri, Y.~Sendouda, and J.~Yokoyama, {\it {New cosmological
  constraints on primordial black holes}},  {\em Phys.Rev.} {\bf D81} (2010)
  104019, [\href{http://arxiv.org/abs/0912.5297}{{\tt arXiv:0912.5297}}].

\bibitem{Lin:2012gs}
C.-M. Lin and K.-W. Ng, {\it {Primordial Black Holes from Passive Density
  Fluctuations}},  {\em Phys.Lett.} {\bf B718} (2013) 1181--1185,
  [\href{http://arxiv.org/abs/1206.1685}{{\tt arXiv:1206.1685}}].

\bibitem{Alabidi:2012ex}
L.~Alabidi, K.~Kohri, M.~Sasaki, and Y.~Sendouda, {\it {Observable Spectra of
  Induced Gravitational Waves from Inflation}},  {\em JCAP} {\bf 1209} (2012)
  017, [\href{http://arxiv.org/abs/1203.4663}{{\tt arXiv:1203.4663}}].

\bibitem{Agashe:2014kda}
{\bf Particle Data Group} Collaboration, K.~Olive et~al., {\it {Review of
  Particle Physics}},  {\em Chin.Phys.} {\bf C38} (2014) 090001.

\bibitem{Kofman:1997yn}
L.~Kofman, A.~D. Linde, and A.~A. Starobinsky, {\it {Towards the theory of
  reheating after inflation}},  {\em Phys.Rev.} {\bf D56} (1997) 3258--3295,
  [\href{http://arxiv.org/abs/hep-ph/9704452}{{\tt hep-ph/9704452}}].

\bibitem{1978amms.book.....B}
C.~M. {Bender} and S.~A. {Orszag}, {\em {Advanced Mathematical Methods for
  Scientists and Engineers}}.
\newblock 1978.

\bibitem{Martin:2002vn}
J.~Martin and D.~J. Schwarz, {\it {WKB approximation for inflationary
  cosmological perturbations}},  {\em Phys.Rev.} {\bf D67} (2003) 083512,
  [\href{http://arxiv.org/abs/astro-ph/0210090}{{\tt astro-ph/0210090}}].

\bibitem{Felder:2000hq}
G.~N. Felder and I.~Tkachev, {\it {LATTICEEASY: A Program for lattice
  simulations of scalar fields in an expanding universe}},  {\em
  Comput.Phys.Commun.} {\bf 178} (2008) 929--932,
  [\href{http://arxiv.org/abs/hep-ph/0011159}{{\tt hep-ph/0011159}}].

\bibitem{Frolov:2008hy}
A.~V. Frolov, {\it {DEFROST: A New Code for Simulating Preheating after
  Inflation}},  {\em JCAP} {\bf 0811} (2008) 009,
  [\href{http://arxiv.org/abs/0809.4904}{{\tt arXiv:0809.4904}}].

\bibitem{Easther:2010qz}
R.~Easther, H.~Finkel, and N.~Roth, {\it {PSpectRe: A Pseudo-Spectral Code for
  (P)reheating}},  {\em JCAP} {\bf 1010} (2010) 025,
  [\href{http://arxiv.org/abs/1005.1921}{{\tt arXiv:1005.1921}}].

\bibitem{Huang:2011gf}
Z.~Huang, {\it {The Art of Lattice and Gravity Waves from Preheating}},  {\em
  Phys.Rev.} {\bf D83} (2011) 123509,
  [\href{http://arxiv.org/abs/1102.0227}{{\tt arXiv:1102.0227}}].

\bibitem{GABE}
``http://cosmo.kenyon.edu/gabe.html.''

\bibitem{Amin:2010dc}
M.~A. Amin, R.~Easther, and H.~Finkel, {\it {Inflaton Fragmentation and
  Oscillon Formation in Three Dimensions}},  {\em JCAP} {\bf 1012} (2010) 001,
  [\href{http://arxiv.org/abs/1009.2505}{{\tt arXiv:1009.2505}}].

\bibitem{Amin:2010xe}
M.~A. Amin, {\it {Inflaton fragmentation: Emergence of pseudo-stable inflaton
  lumps (oscillons) after inflation}},
  \href{http://arxiv.org/abs/1006.3075}{{\tt arXiv:1006.3075}}.

\bibitem{Lozanov:2014zfa}
K.~D. Lozanov and M.~A. Amin, {\it {End of inflation, oscillons, and
  matter-antimatter asymmetry}},  {\em Phys.Rev.} {\bf D90} (2014), no.~8
  083528, [\href{http://arxiv.org/abs/1408.1811}{{\tt arXiv:1408.1811}}].

\bibitem{Sfakianakis:2012bq}
E.~I. Sfakianakis, {\it {Analysis of Oscillons in the SU(2) Gauged Higgs
  Model}},  \href{http://arxiv.org/abs/1210.7568}{{\tt arXiv:1210.7568}}.

\bibitem{Graham:2007ds}
N.~Graham, {\it {Numerical Simulation of an Electroweak Oscillon}},  {\em
  Phys.Rev.} {\bf D76} (2007) 085017,
  [\href{http://arxiv.org/abs/0706.4125}{{\tt arXiv:0706.4125}}].

\bibitem{Ade:2015lrj}
{\bf Planck Collaboration} Collaboration, P.~Ade et~al., {\it {Planck 2015. XX.
  Constraints on inflation}},  \href{http://arxiv.org/abs/1502.02114}{{\tt
  arXiv:1502.02114}}.

\end{thebibliography}\endgroup

\end{document}